\documentclass[useAMS,usenatbib]{mn2e}
\usepackage{times}
\usepackage{amssymb}
\usepackage{graphicx}
\usepackage{rotating}
\usepackage{lscape}
\usepackage{multirow}

\begin{document}

\title[Red sequence slope evolution]{The evolution of the red sequence slope in massive galaxy clusters}
\author[J.P.Stott et al.]{J.\,P. Stott$^{1, 2}$\thanks{E-mail: jps@astro.livjm.ac.uk}, K.\,A. Pimbblet$^3$, A.\,C. Edge$^2$, G.\,P. Smith$^{4, 5}$, J.\, L. Wardlow$^2$\\
\\
$^1$Astrophysics Research Institute, Liverpool John Moores University, Twelve Quays House, Egerton Wharf, Birkenhead, CH41 1LD, UK\\
$^2$Institute for Computational Cosmology, Department of Physics, University of Durham, South Road, Durham DH1 3LE, UK\\
 $^3$Department of Physics, University of Queensland, Brisbane, QLD 4072, Australia\\
 $^4$California Institute of Technology, Mail Code 105-24,
  Pasadena, CA 91125, USA\\
 $^5$  School of Physics $\&$ Astronomy, University of Birmingham, Edgbaston, Birmingham, B15 2TT, UK \\
}

\date{}

\pagerange{\pageref{firstpage}--\pageref{lastpage}} \pubyear{2008}

\maketitle

\label{firstpage}

\begin{abstract}

\noindent We investigate the evolution of the optical and near-infrared colour-magnitude relation in an homogeneous sample of massive clusters from $z=1$ to the present epoch. By comparing deep {\it Hubble Space Telescope} ACS imaging of X-ray selected MACS survey clusters at $z\sim0.5$ to the similarly selected LARCS sample at $z\sim0.1$ we find that the rest-frame $\delta(U - V)/\delta V$ slope of the colour-magnitude relation evolves with redshift which we attribute to the build up of the red sequence over time. This rest frame slope evolution is not adequately reproduced by that predicted from semi-analytic models based on the Millennium Simulation despite a prescription for the build up of the red sequence by in-falling galaxies, `strangulation'. We observe no strong correlation between this slope and the cluster environment at a given redshift demonstrating that the observed evolution is not due to a secondary correlation. Also presented are near-infrared UKIRT WFCAM observations of the LARCS clusters which confirm and improve on the the result from \cite{Stott2007a} finding that there has been a two-fold increase in faint $M_V>-20$ galaxies on the red sequence since $z=0.5$ to a significance of 5$\sigma$.
\end{abstract}


\section{Introduction}
\label{sec:introcm}

Galaxy clusters are important laboratories for the study of galaxy formation and evolution as they contain a concentrated population of diverse galaxies in a relatively small volume. Early workers in the field found that if a colour-magnitude diagram is plotted for members of local clusters such as Virgo and Coma, the Elliptical/S0 galaxies were found to be confined to a prominent linear feature in colour space \citep{vis1977}. This feature is known as the red sequence and has a very small intrinsic scatter (typically $<$0.1 mag) which has been interpreted as evidence that the passive galaxies in clusters formed coevally at high redshift \citep{Bower1992}.

The red sequence is observed to have a slope such that the faint galaxies are bluer than the bright cluster members. The origin of the slope has been controversial due to the age metallicity degeneracy for stellar populations. \cite{worthey1995} showed that the sequence of colours that comprise the slope can be equally well explained by a progressive decrease in either metallicity or stellar age. The slope is now thought to be due to a mass-metallicity relation along the red sequence as this degeneracy was broken by comparing colour-magnitude simulations to observations of distant clusters \citep{Kodama1997}. The origin of the mass-metallicity relation is thought to be the heating of the interstellar medium (ISM) by supernovae which triggers the formation of a galactic wind when the thermal energy of the gas exceeds the binding energy. This wind ejects gas more efficiently in smaller galaxies due to their shallower potential wells, resulting in the trend of increased metallicity with mass which manifests itself as the massive metal-rich galaxies appearing progressively redder than their less massive counterparts
(\citealt{carlberg1984}; \citealt{arimoto1987}). Within the hierarchical picture of galaxy evolution, red sequence galaxies form from the mergers of star forming disc galaxies with the largest ellipticals being the product of the merger of the largest disc systems and are thus the most metal rich \citep{Kauffmann1998}. 

The observed frame colour-magnitude relations for clusters at similar redshifts are found to have comparable slopes which have been shown to change with redshift (\citealt{Lopez1997}; \citealt{Glad1998};  \citealt{Lopez2004}). One explanation for this change in the observed slope is that it is a result of $K$ correction and an evolution in the mass metallicity relation in cluster ellipticals (\citealt{Kodama1997}). The change of slope with redshift has been used as a method to constrain cluster galaxy evolution, with comparisons of the observed slope evolution to models suggesting that the elliptical galaxies in cluster cores have been in place since at least $z=2$ \citep{Glad1998}. However other studies of the rest frame slope have found little or no evolution suggesting that $K$ correction is the dominant factor (\citealt{stanford98}; \citealt{blakes2003}; \citealt{mei2006}). This non-evolution has been used as evidence to favour monolithic collapse over hierarchical methods of galaxy formation. 

Current research on the build up of the red sequence suggests there may indeed be an age contribution to the red sequence slope and its evolution as faint, sub-L$^{*}$ galaxies, thought to have undergone recent star formation, are found to be transforming onto the sequence over time as they fall into the dense cluster environment (\citealt{delucia2007dgr}; \citealt{Stott2007a}; \citealt{smith08}). This is thought to be connected to the observation that there is a greater abundance of S0 galaxies in local clusters compared to their high redshift counterparts, whose progenitors may be the quenched star forming galaxies in higher redshift systems (\citealt{dres1997}; \citealt{smail1998}).  If this transformation scenario is correct then we expect to see evidence for it in the evolution of the rest frame colour-magnitude relation. 

Regardless of its origin, empirical observations and models of the colour and slope of the observed red sequence can be used in combination as a template to predict the redshift of a cluster. This is found to be in good agreement with spectroscopic redshifts and can therefore be used as an effective two waveband photometric redshift. This technique is important for current and future large area photometric surveys which hope to study cluster abundance (e.g. RCS, \citealt{Glad2000}; Pan-STARSS; SDSS, \citealt{SDSS}; UKIDSS, \citealt{Law2007}, \citealt{Swin2007}).

In this work we will investigate the red sequence slope evolution in observed optical ($\delta(V - I)/\delta I$) and near infrared ($\delta(J - K)/\delta K $) bands and rest frame optical ($\delta(U - V)/\delta V$) for a homogeneous sample of X-ray selected galaxy clusters in the range $z\sim$0--1. This is the most comprehensive study of the red sequence slope undertaken thus far. We compare our findings with latest synthetic slopes calculated from analysis of the semi-analytical model of \cite{Bower2006} based on the Millennium N-body simulation \citep{Springel2005}. This model includes feedback from active galactic nuclei (AGN) which quenches star formation in massive halos to match the observed break in the luminosity function seen at bright magnitudes. Another important process in this model is `strangulation' which describes the stripping of a galaxy's hot gas halo as it falls into a cluster leading to a cessation in star formation in lower mass galaxies thus providing a mechanism for the build up of passive red galaxies in the cluster \citep{larson1980}.

\noindent Lambda Cold Dark Matter ($\Lambda$CDM) cosmology ($\Omega_{M}=$0.3, $\Omega_{Vac}=$0.7, $H_{0}=$70 km s$^{-1}$ Mpc$^{-1}$) is used throughout this work.

\section{Observations and Reduction}

\subsection{Cluster sample}
\label{sec:sam}
To observe the evolution of the red sequence we study X-ray selected clusters in the range $z=$0--1 in both near-infrared and optical bands. This is a large sample of the most X-ray luminous clusters known (L$_{X}>10^{44} $erg s$^{-1}$, 0.1 -- 2.4 keV) which, with the exception of a handful of additional archival clusters, are all sourced from the \emph{ROentgen SATellite} (\emph{ROSAT}) All Sky Survey, although belong to the various sub-samples named below. Such massive clusters are ideal for this study as their colour-magnitude relations are well populated. The motivation for studying an X-ray selected sample of clusters is to ensure that we are observing objects in similar high mass, high density environments. This homogeneity is key to our study as we wish to compare clusters over a range of redshifts. 

The main cluster samples studied in this paper, the MAssive Cluster Survey (MACS, \citealt{MACS2001}; \citealt{Ebeling2007}) and the Las Campanas/AAT Rich Cluster Survey (LARCS, \citealt{LARCS2001}, 2006), are sufficiently X-ray luminous that they should correspond to the most extreme environments at their respective epochs. The median X-ray luminosities of the high and low redshift samples are 16.1 and 7.4$\times10^{44}$ erg s$^{-1}$, respectively, corresponding to a difference of less than a factor of 2 in the typical total mass \citep{popesso2005}. However, an important issue to address is that the mass of the z$\sim$0.5 progenitors of the LARCS clusters may be even lower than the MACS clusters. If we include the growth in halo mass through N-body mergers from \cite{Bower2006}, we see that the progenitors of the LARCS clusters at $z\sim$0.5 could be up to 3 times less massive than the MACS sample (with corresponding X-ray luminosities of 2 $\times10^{44}$ erg s$^{-1}$, \citealt{popesso2005}). There is no evidence for strong variations in the galaxy luminosity function between clusters spanning such a relatively modest difference in typical mass (de Propris et al. 1999). We therefore expect that any differences between the galaxy populations in these two samples will primarily reflect evolutionary differences between $z\sim$0.5 and $z\sim$0.1. 

The 25 clusters observed in $V$ and $I$ or $B$ and $R$ bands belong to the following surveys:  MACS ($V_{F555W}$ and $V_{F814W}$, Cycle 12 GO project 9722); LARCS ($B$ and $R$) and archival {\it Hubble Space Telescope} ({\it HST}) data ($V_{F555W/F606W}$ and $V_{F814W}$). The observations were taken with the instruments: the Advanced Camera for Surveys (ACS) and the Wide Field Planetary Camera (WFPC) on {\it HST} and the 1m Swope telescope at Las Campanas Observatory. For a more detailed description of the data and reduction see  \cite{Stott2007a} (MACS) and \cite{LARCS2001} (LARCS).

The 35 clusters studied in the near-infrared ($J$ and $K$ band) belong to: The MACS survey \citep{MACS2001}, the LARCS survey \citep{LARCS2001}, the \emph{ROSAT} Brightest Cluster Survey (BCS) and extended BCS ~\citep{Ebeling2000} and the X-ray Brightest Abell Clusters Survey (XBACS, \citealt{xbacs1996}). The observations were taken with: the Wide field InfraRed Camera (WIRC) instrument on the Palomar 200" Hale telescope the Infrared Spectrometer And Array Camera (ISAAC) on the Very Large Telescope (VLT) and the Wide Field CAMera (WFCAM) on United Kingdom Infrared Telescope (UKIRT). An additional low redshift data point is included for the Coma Cluster which we sourced from a combination of the Two Micron All Sky Survey (2MASS) extended and point source catalogues ~\citep{2mass2006}. We refer the reader to \cite{stott2007b} for a more detailed description of the data and their reduction.

We include a high redshift cluster (ClJ1226.9+3332, $z\sim0.9$) from the Wide Angle \emph{ROSAT} Pointed Surveys (WARPS, \citealt{WARPS1997}; \citealt{WARPS1998}). This X-ray selected cluster was observed in the J and K bands using the UKIRT Fast-Track Imager (UFTI) camera on UKIRT \citep{Ellis2004}. 

\subsection{Photometry}
\label{sec:phot}
The colour ($V - I$, $B - R$ and $J - K$) photometry extracted for the cluster members employs 9kpc diameter apertures for colours and the magnitude used is {\sc SExtractor}'s `Best' magnitude ~\citep{sextractor1996}. This choice of aperture is significantly greater than the seeing conditions in which the low redshift ground-based optical and near-infrared data were taken, typically $\sim$1.0 -- 1.2 arcsec, giving LARCS sample photometric apertures of $\sim$4" whereas the higher redshift MACS sample is extracted with $\sim$1.4" apertures, considerably larger than the $\it HST$ ACS PSF. We are therefore confident that we are collecting a comparable fraction of light for galaxies at different redshifts. We observe no trend between colour and seeing for our LARCS low redshift optical data, although the seeing is consistent, and as the high redshift sample is observed with {\it HST} this is not a concern. We also investigate this for the $z\sim0.1 - 0.3$ clusters observed in the near-infrared wavebands for which colour is again independent of seeing, even though the conditions are more variable for our additional {\it ROSAT} clusters sample (0.9" -- 1.5"). 

The {\sc SExtractor} colour photometry was run in dual mode, on PSF matched images using the {\sc iraf psfmatch} package, with the 9kpc `red'-band ($R$, $I$, $K$) apertures used to extract the corresponding `blue'-band ($B$, $V$, $J$) photometry. This is to ensure the same size aperture for both bands which is important for good colour determination. Star and Galaxy separation is performed using {\sc SExtractor} where detected objects with {\sc class\_star}$<$0.1 and/or $J-K>$0.95 (for the near-infrared sample) are classified as galaxies with the stars removed from the analysis. We only consider galaxies within 600kpc radius of the cluster centre, to limit contamination from field galaxies. All of our observations reach a depth which allows us to see to at least 4 magnitudes fainter than the BCG and thus perform a reliable fit to the slope and probe the sub-L$^{\star}$ galaxy population where the red sequence build up is taking place.  

A potential problem with all aperture photometry is that the galaxies themselves may have significant internal colour gradients thought to be associated with a metallicity gradient between the outer and inner parts of the system. As stated above we are observing the same fraction of light from galaxies at all redshifts which should limit this effect unless there is a significant evolution in the colour gradient of galaxies with redshift, although this would be an interesting effect in itself. Studies of the internal colour gradients of moderate redshift cluster galaxies using high quality {\it HST} data have found no such evolution in mean colour gradient and redshift, with the individual cluster galaxies appearing to have random colour gradients \cite{tamura2000}. We therefore conclude that any variations in internal colour gradient would act randomly to make galaxies appear either redder or bluer resulting in increased scatter of the red sequence but not an evolution in its slope.

\begin{table*}
\begin{center}
\caption[]{Details of the cluster sample used in our analysis which, with the exception of a handful of additional clusters, belong to \emph{ROSAT} All Sky Survey. The LARCS and MACS redshifts are from \cite{LARCS2006} and \cite{Ebeling2007} respectively. $\kappa_{opt.}$ and $\kappa_{JK}$ are the observed optical and near-infrared slopes.}
\label{tab:samplecm}
\small\begin{tabular}{lcccccc}
\hline
Cluster & R.A.\ & Dec.\ & $z$ & $L_X$ & $\kappa_{opt.}$ & $\kappa_{JK}$\\
&\multicolumn{2}{c}{(J2000)}&&($10^{44}$erg\,s$^{-1}$)\\
\hline
%\hline
\noalign{\medskip}
\noalign{\smallskip}
\multispan{2}{LARCS $z\sim 0.1$ Sample \hfil}\\
\noalign{\smallskip}
%===========================================
Abell\,22&00 20 38.64&$-$25 43 19& 0.142&5.3   & -0.072 $\pm$0.014 &-0.023$\pm$0.010\\
Abell\,550&05 52 51.84&$-$21 03 54& 0.099&7.1  & -0.055 $\pm$0.013 &-0.024$\pm$0.011\\ 
Abell\,1084&10 44 30.72&$-$07 05 02& 0.132&7.4  & -0.058 $\pm$0.013 &-0.033$\pm$0.007\\
Abell\,1285&11 30 20.64&$-$14 34 30& 0.106&5.45 &-0.036 $\pm$0.020 &...\\
Abell\,1437&12 00 25.44&$+$03 21 04& 0.134&7.7  &-0.037 $\pm$0.017 &-0.010$\pm$0.009\\
Abell\,1650&12 58 41.76&$-$01 45 22& 0.084&7.8  & -0.047 $\pm$0.011 &-0.019$\pm$0.009\\
Abell\,1651&12 59 24.00&$-$04 11 20& 0.085&8.3  & -0.048 $\pm$0.009 &-0.002$\pm$0.007\\
Abell\,1664&13 03 44.16&$-$24 15 22& 0.128&5.34 & -0.054 $\pm$0.011 &-0.021$\pm$0.010\\
Abell\,2055&15 18 41.28&$+$06 12 40& 0.102&4.8  & -0.066 $\pm$0.011 &-0.021$\pm$0.009\\
Abell\,3888&22 34 32.88&$-$37 43 59& 0.153&14.5 & -0.047 $\pm$0.022 &...\\
%===========================================
\noalign{\medskip}
\multispan{2}{MACS $z= 0.4$--$0.7$ Sample \hfil}\\
\noalign{\smallskip}
%===========================================
MACS\,J0018.5$+$1626 & 00:18:33.68 &$+$16:26:15 & 0.541 & 18.74 &...&-0.048$\pm$0.009\\
MACS\,J0025.4$-$1222 &00 25 15.84 &$-$12 19 44 & 0.478 &12.4 & -0.106 $\pm$0.021 &-0.047$\pm$0.010\\
MACS\,J0257.6$-$2209 &02 57 07.96 &$-$23 26 08 & 0.504 &15.4  & -0.105 $\pm$0.014 &-0.065$\pm$0.012 \\
MACS\,J0454.1$-$0300 & 04:54:11.13 &$-$03:00:53.8 & 0.550 & 16.86 &...&-0.040 $\pm$0.013\\
MACS\,J0647.7$+$7015 &06 47 51.45 &$+$70 15 04 & 0.584 &21.7  &-0.090 $\pm$0.024 &...\\
MACS\,J0712.3$+$5931 & 07 12 20.45 &$+$59 32 20  & 0.328 &6.8    &-0.044 $\pm$0.006 &...\\
MACS\,J0717.5$+$3745 &07 17 31.83 &$+$37 45 05 & 0.548 &27.4  &-0.068 $\pm$0.012 &...\\
MACS\,J0744.8$+$3927 &07 44 51.98 &$+$39 27 35 & 0.686 &25.9  &-0.081 $\pm$0.023 &...\\
MACS\,J0911.2$+$1746&09 11 10.23 &$+$17 46 38 & 0.506 &13.2  &-0.087 $\pm$0.011 &...\\
MACS\,J1149.5$+$2223 &11 49 34.81 &$+$22 24 13 & 0.544 &17.3  &-0.101 $\pm$0.011 &...\\
MACS\,J1354.6$+$7715 & 13 54 19.71 &$+$77 15 26  & 0.397 &8.2    &-0.057 $\pm$0.018 &...\\
MACS\,J1359.8$+$6231 & 13:59:54.32 & $+$62:30:36.3 & 0.330 &  8.83 &...&-0.039 $\pm$0.007\\
MACS\,J1423.8$+$2404 &14 23 47.95 &$+$24 04 59 & 0.544 &15.0  &-0.078 $\pm$0.013 &...\\
MACS\,J2129.4$-$0741 &21 29 25.38 &$-$07 41 26 & 0.570 &16.4  &-0.115 $\pm$0.017 &-0.068$\pm$0.009\\
MACS\,J2214.9$-$1359 &22 14 56.51 &$-$14 00 17 & 0.495 &17.0  &-0.100 $\pm$0.012 &-0.035$\pm$0.014\\
%===========================================
\noalign{\medskip}
\multispan{2}{Additional \emph{ROSAT} All Sky Survey clusters \hfil}\\
\noalign{\smallskip}
Abell\,115 & 00:56:00.24 & $+$26:20:31.7 & 0.197  & 14.59 &...&-0.031 $\pm$0.006\\
Abell\,209 & 01:31:52.51 & $-$13:36:41.0 & 0.209  & 13.75 &...&-0.022 $\pm$0.009\\
Abell\,291 & 02:01:46.80 & $-$02:11:56.9 & 0.196 & 4.24 &...&-0.024 $\pm$0.009\\
Abell\,665 & 08:30:57.34 & $+$65:50:31.4 & 0.182 & 16.33 &...&-0.020 $\pm$0.008\\
Abell\,773 & 09:17:53.57 & $+$51:44:02.5 & 0.217 & 13.08 &...&-0.033 $\pm$0.006\\
Abell\,1201 & 11:12:54.50 & $+$13:26:08.9 & 0.169 & 6.28 &...&-0.013 $\pm$0.007\\
Abell\,1246 & 11:23:58.75 & $+$21:28:47.3 & 0.190 & 7.62 &...& -0.031 $\pm$0.008\\
Abell\,1703  & 13 15 00.70 &$+$51 49 10  & 0.258 &8.7   &-0.057 $\pm$0.012 &...\\
Abell\,1758 & 13:32:38.59 & $+$50:33:38.7 & 0.279 & 11.68&...&-0.027 $\pm$0.011\\
Abell\,1763 & 13:35:20.14 & $+$41:00:03.8 & 0.223 & 14.93&...& -0.029 $\pm$0.010\\
Abell\,1914 & 14:25:56.64 & $+$37:48:59.4 & 0.171 & 18.39 &...&-0.021 $\pm$0.006\\
Abell\,2111 & 15:39:41.81 & $+$34:24:43.3 & 0.229 & 10.94 &...&-0.033 $\pm$0.012\\
Abell\,2163 & 16:15:33.57 & $-$06:09:16.8 & 0.203 & 37.50 &...&-0.022 $\pm$0.006\\
Abell\,2218 & 16:35:49.39 & $+$66:12:45.1 & 0.176 & 9.30 &...&-0.025 $\pm$0.007\\
Abell\,2445 & 22:26:55.80 & $+$25:50:09.4 & 0.165 & 4.00 &...&-0.027 $\pm$0.009\\
RX\,J1720.1$+$2638 & 17:20:10.08 & $+$26:37:33.5 & 0.164 & 6.66 &...& -0.023 $\pm$0.007\\
Zw\,1432 & 07:51:25.15 & $+$17:30:51.8 & 0.186 & 5.27 &...& -0.022 $\pm$0.010\\
%===========================================
\noalign{\medskip}
\multispan{2}{Additional archival clusters \hfil}\\
\noalign{\smallskip}
Cl\,J0152$-$1357     & 01 52 43.91 &$-$13 57 21  & 0.831 &5.0   &-0.084 $\pm$0.026 &...\\
Cl\,J1226.9$+$3332   & 12 26 58.13 &$+$33 32 49  & 0.890 &20.0  &-0.088 $\pm$0.016 &-0.088$\pm$0.010\\
Coma Cluster&12:59:48.70& $+$27:58:50.0 & 0.0231 &7.26  &-0.075 $\pm$0.015& -0.017 $\pm$0.009\\
MS1054$-$0321 & 10:57:00.20 & $-$03:37:27.4 & 0.830 & 23.30 &...& -0.073 $\pm$0.011\\
RCS0224$-$0002 & 02:24:00.00 & $-$0:02:00.0 & 0.770 & 0.70 &...& -0.055 $\pm$0.020\\
%===========================================
\hline
\end{tabular}
\end{center}
\end{table*}
\normalsize

\label{tab:samplecm}

\subsection{WFCAM data}
\label{sec:wfcam}
We obtained near-infrared $J$ and $K$ band data for 8 of the LARCS $z\sim0.1$ clusters with WFCAM on UKIRT. WFCAM consists of 4 detectors in a square, each separated by a gap comparable in size to a single detector, with a central autoguider. Each detector is a Rockwell Hawaii II 2048 $\times$ 2048 PACE HgCdTe array, with pixel size 0.4 arcsec. WFCAM single pointings or `footprints' do not observe a contiguous area of sky so to create a mosaiced image four of these WFCAM footprints are tiled to make a 4 detector $\times$ 4 detector image. In addition to this tiling the WFCAM images incorporate `microstepping' which is a small dither pattern in the observation. The purpose of this process is to improve the point spread function sampling of the observations as the 0.4$''$ pixel size of WFCAM is comparable to the best atmospheric seeing. The microstepping is performed by observing 4 images each offset from each other by a whole number of pixels and a 1/2 pixel in a 2 $\times$ 2 grid pattern. These images are then coadded together with the fractional offsets taken into account so that the resulting image has double the spatial sampling. Therefore, a detector image with an original resolution of 0.4 $''$/pixel and size 2048 $\times$ 2048 pixels are converted to a higher resolution 0.2 $''$/pixel 4096 $\times$ 4096 image. 

The observations took place during service mode on the nights of 5th December 2006 and 19th of April 2007 in 1.0" and 1.2" seeing conditions respectively. The total integration times for the $J$ and $K$ band images are 200s each, composed of 10s exposures in a 5 point dither pattern with 2$\times$2 microstepping. To create a contiguous Mosaic of WFCAM images we run the {\sc Terapix SWarp} software on the data which gives an image with a world coordinate system consistent to within 0.1" of the 2MASS and the United States Naval Observatory (USNO) star catalogue across the whole field of view. To extract the $J$ and $K$ band catalogues from the mosaiced images we use the {\sc SExtractor} software in dual mode so that the $K$ band catalogue detections are used to extract the $J$ band photometry in an identical process to above. The photometry is calibrated using the 2MASS point source catalogue and we find the $J$ and $K$ band 5 sigma vega limits of these data are 19.50 and 17.75 magnitudes respectively.

\subsubsection{Quantifying the dwarf to giant ratio of the LARCS clusters}

\begin{figure}
\centering
\includegraphics[width=0.5\textwidth]{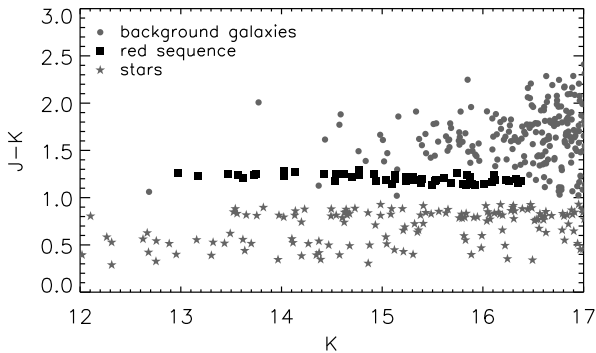}
\caption[]{The $(J - K)$ vs $K$ colour-magnitude diagram for the cluster Abell 1084 which demonstrates the ease at which near-infrared colours separate the red sequence from stellar and field galaxy contamination}
\label{fig:cmwfcam}
\end{figure}

The WFCAM instrument allows us to create contiguous images that are $\sim$0.9 degrees on the side which corresponds to 5.6Mpc at $z=0.1$. With images of this size we can study both the cluster and the surrounding field galaxy population. A number of recent papers have studied the form of the colour-magnitude relation in galaxy clusters with some reporting a dearth of faint red sequence galaxies at high redshift (e.g. \citealt{kodama2004}; \citealt{delucia2004dgr}; \citealt{delucia2007dgr}). \cite{Stott2007a} demonstrate that faint red galaxy population in clusters is built up over time by analysing the ratio of faint to luminous galaxies along the red sequence in a homogenous sample of clusters at $z\sim0.1$ and $z\sim0.5$. One of the limiting factors in that study was the uncertainty in the faint end statistical field correction for the LARCS $z\sim0.1$ colour-magnitude relations. However the uncertainty in the field correction is reduced if near infrared observations are employed as it is easier to isolate the cluster red sequence. A comparison of Fig. \ref{fig:cmwfcam} with the equivalent optical colour magnitude diagram, Fig. 1 {(\it bottom right panel)}, of \cite{Stott2007a} illustrates this point. 

By analysing the WFCAM data in an identical way to that described in \cite{Stott2007a} in concert with the UKIDSS Deep eXtragalactic Survey (DXS, Survey Head: Alastair Edge) field observations we can successfully quantify both the cluster and field galaxy population. The field correction is performed by dividing the colour-magnitude space of the cluster and field samples into a two-dimensional histogram and subtracting the field number counts from the cluster population \citep{Stott2007a}. The Poisson errors from this field correction are folded through into the final result. To determine the relative numbers of faint and luminous galaxies on the red sequence we define the red sequence dwarf to giant ratio (RDGR). This quantity is defined as the number of dwarf (faint) galaxies divided by the number of giant (luminous) galaxies on the cluster red sequence after the statistical subtraction of the field galaxy population. The dividing line between dwarfs and giants, for consistency with the \cite{Stott2007a} result, is given in absolute $V$ band magnitude, M$_{V}=-19.9$ and a limiting magnitude of  M$_{V}=-17.75$ for the faintest dwarfs, a correction for passive evolution is applied when studying high redshift samples. We convert these limits to apparent K band magnitudes, $K\sim14.7$ and $K\sim16.8$ respectively (depending on the precise redshift), using a \cite{bruzcharl2003} simple stellar population with solar metallicity and a formation redshift, $z_f=5$. 

\begin{figure}
\centering
\includegraphics[width=0.5\textwidth]{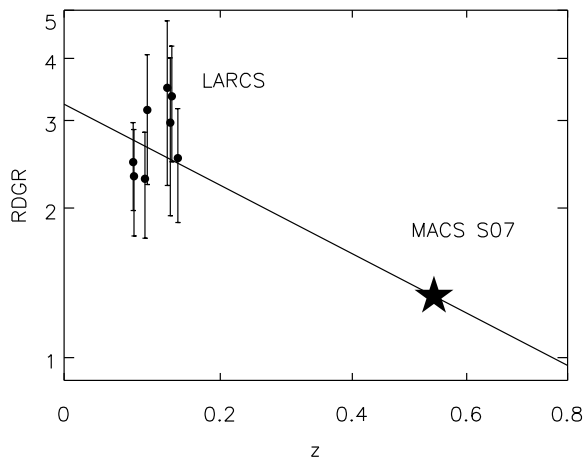}
\caption[]{The evolution of the of the red sequence giant dwarf ratio (RDGR) with redshift comparing the MACS sample from \cite{Stott2007a} with the LARCS WFCAM sample. A fit of the form $(1+z)^{-\beta}$ to the LARCS and MACS samples is plotted which yields $\beta=-2.1\pm0.3$.}
\label{fig:dgr}
\end{figure}

The weighted mean of the RDGRs of the 8 LARCS clusters from this sample is 2.59$\pm$0.24 compared to the \cite{Stott2007a} result of 2.93$\pm$0.45. We can now say that when compared to our $z\sim$0.5 MACS sample, where the DGR=1.33$\pm$0.06, the ratio of dwarf to giant galaxies on the cluster red sequence has increased by a factor of 1.95$\pm$0.20 in the past 5Gyr which confirms and improves the significance of the result of \cite{Stott2007a} from 3$\sigma$ to 5$\sigma$. This is strong evidence for a significant build up of the red sequence in massive clusters. Fig. \ref{fig:dgr} is the equivalent of Fig. 2 {\it (right)} of \cite{Stott2007a} and displays the evolution in the RDGR with redshift for rich galaxy clusters. To parameterise this evolution we fit a $(1+z)^{-\beta}$ power law to the LARCS and MACS samples where $\beta=-2.1\pm0.3$ with all clusters consistent with the fit.

\section{Analysis and Results}

\subsection{Fitting red sequence the slope}
\label{sec:fit}
\begin{figure*}
\centering
\includegraphics[width=0.4\textwidth]{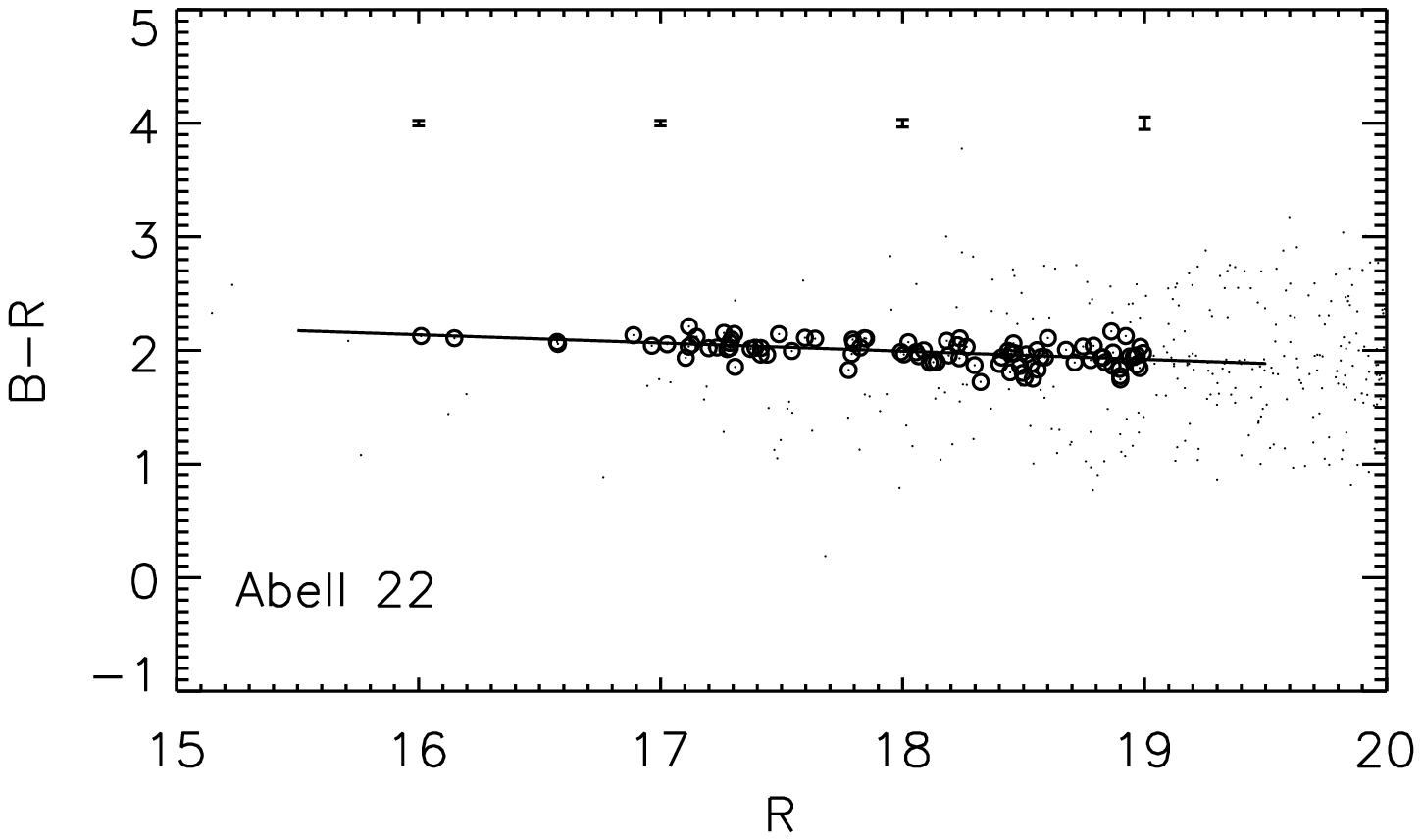}\includegraphics[width=0.4\textwidth]{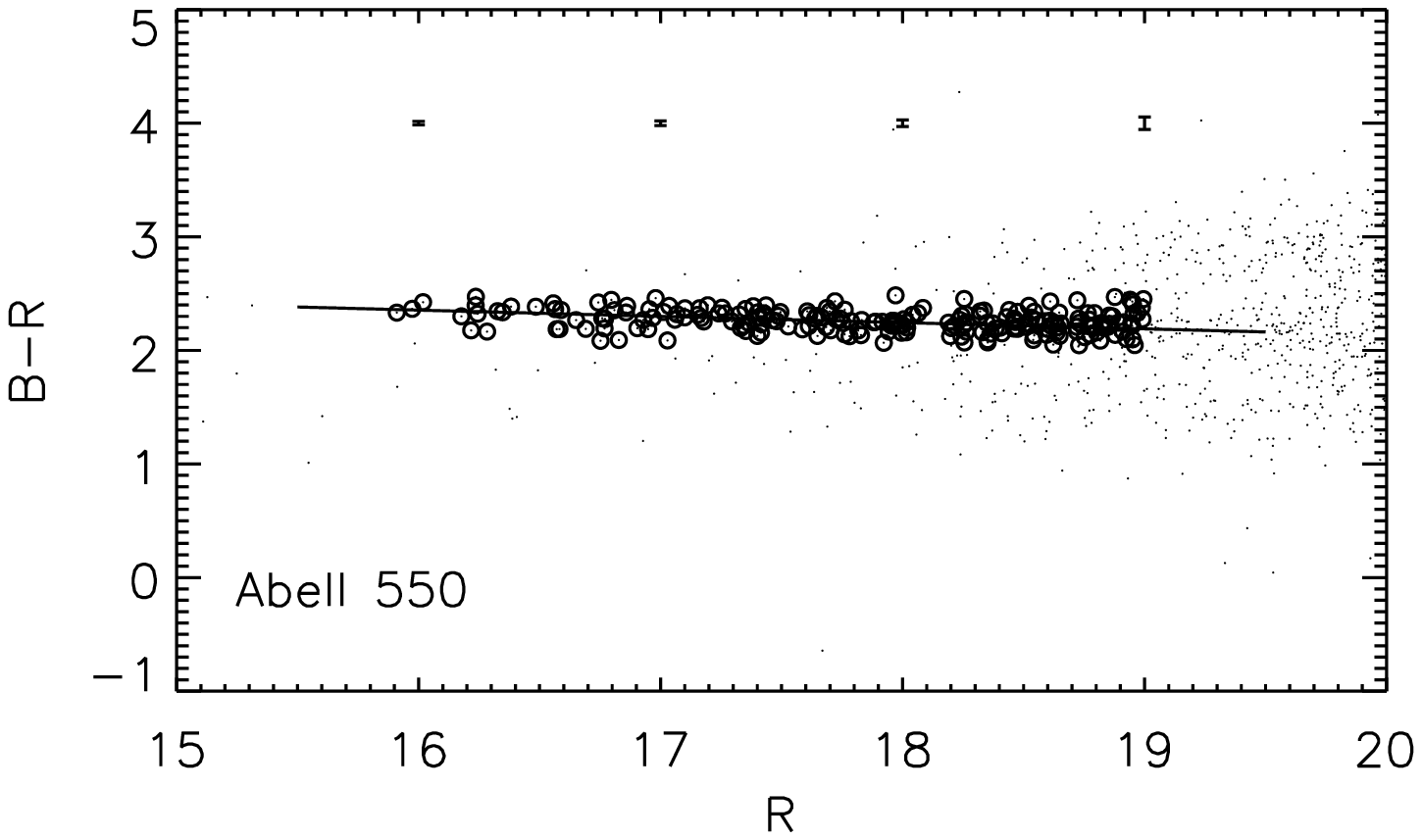}
\includegraphics[width=0.4\textwidth]{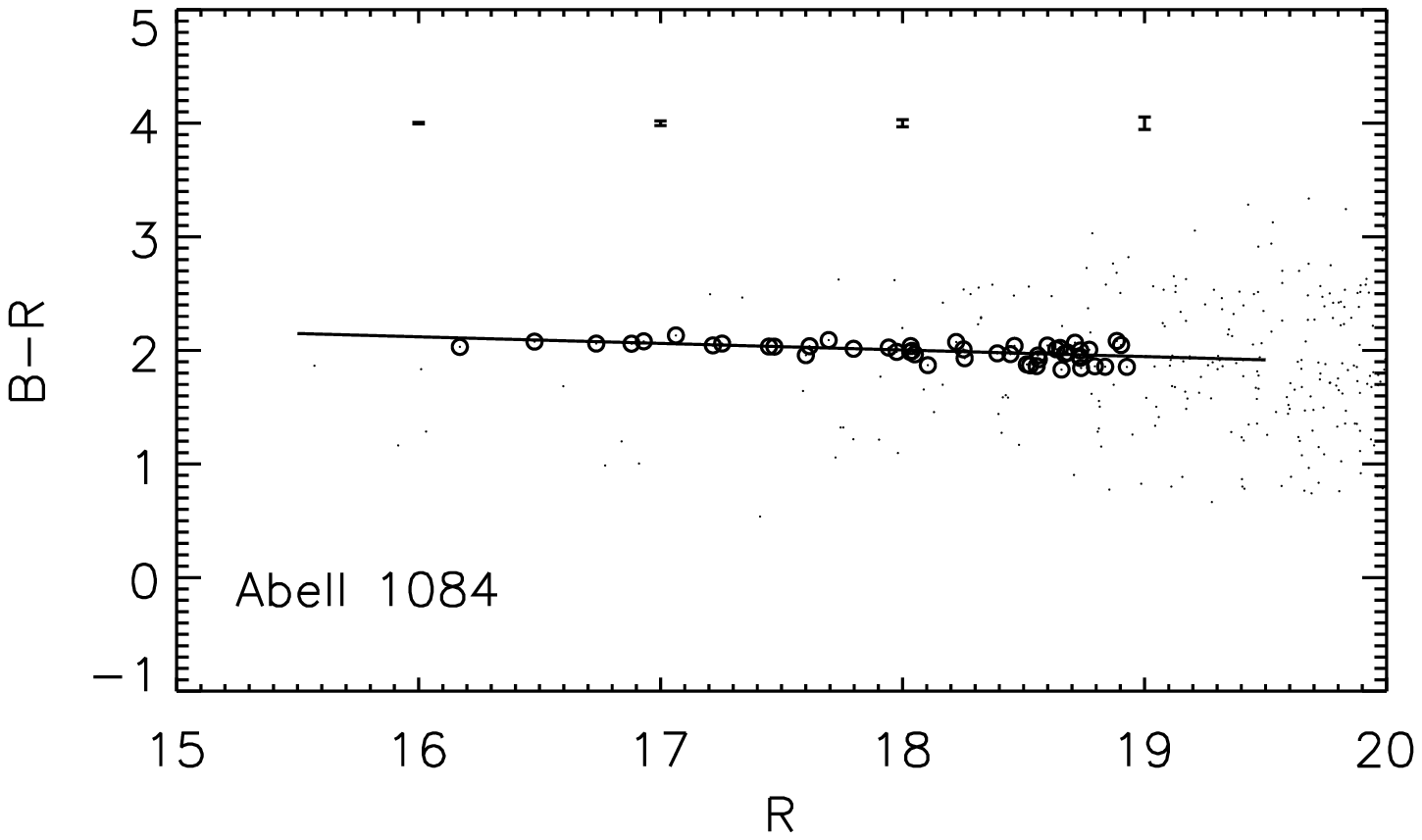}\includegraphics[width=0.4\textwidth]{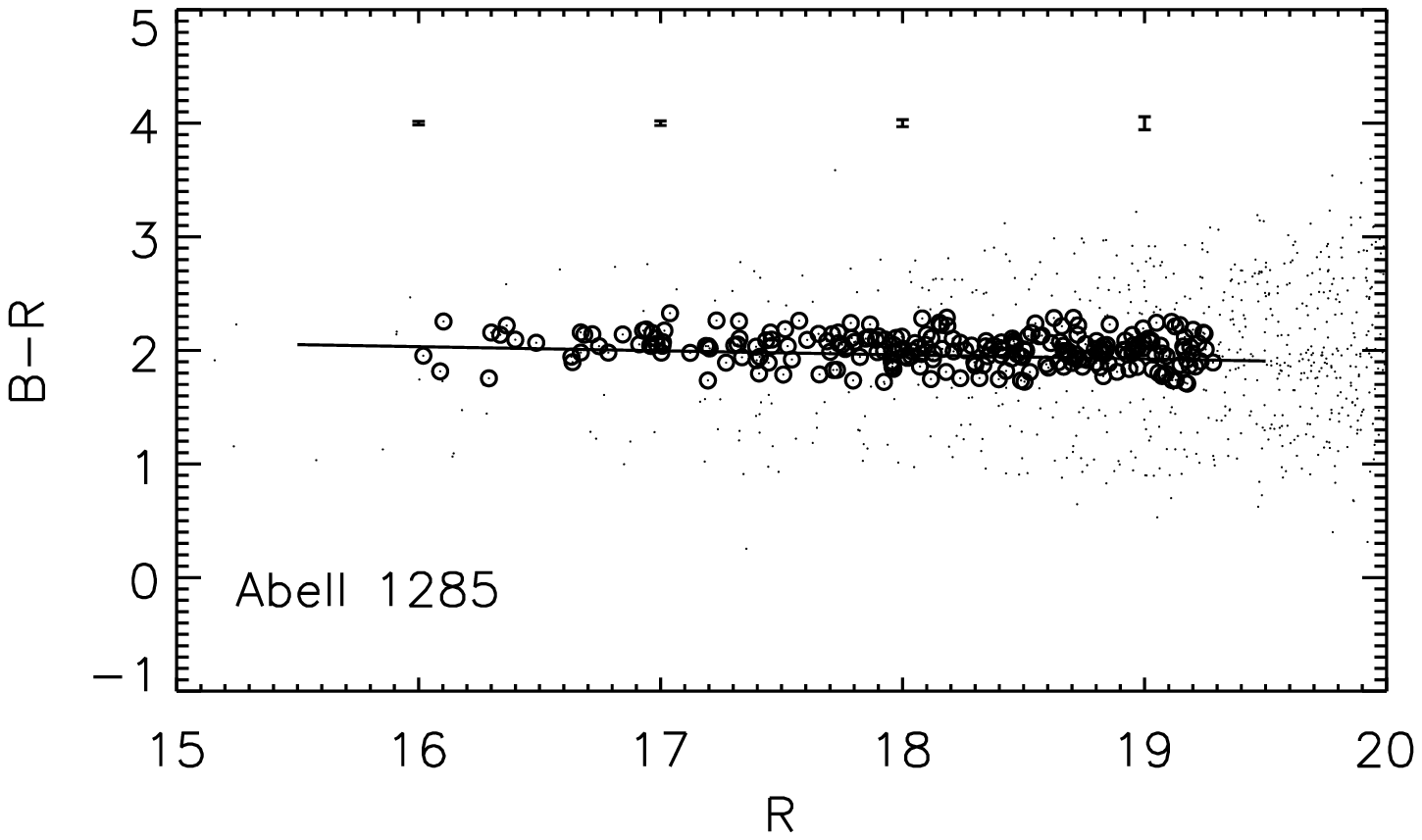}
\includegraphics[width=0.4\textwidth]{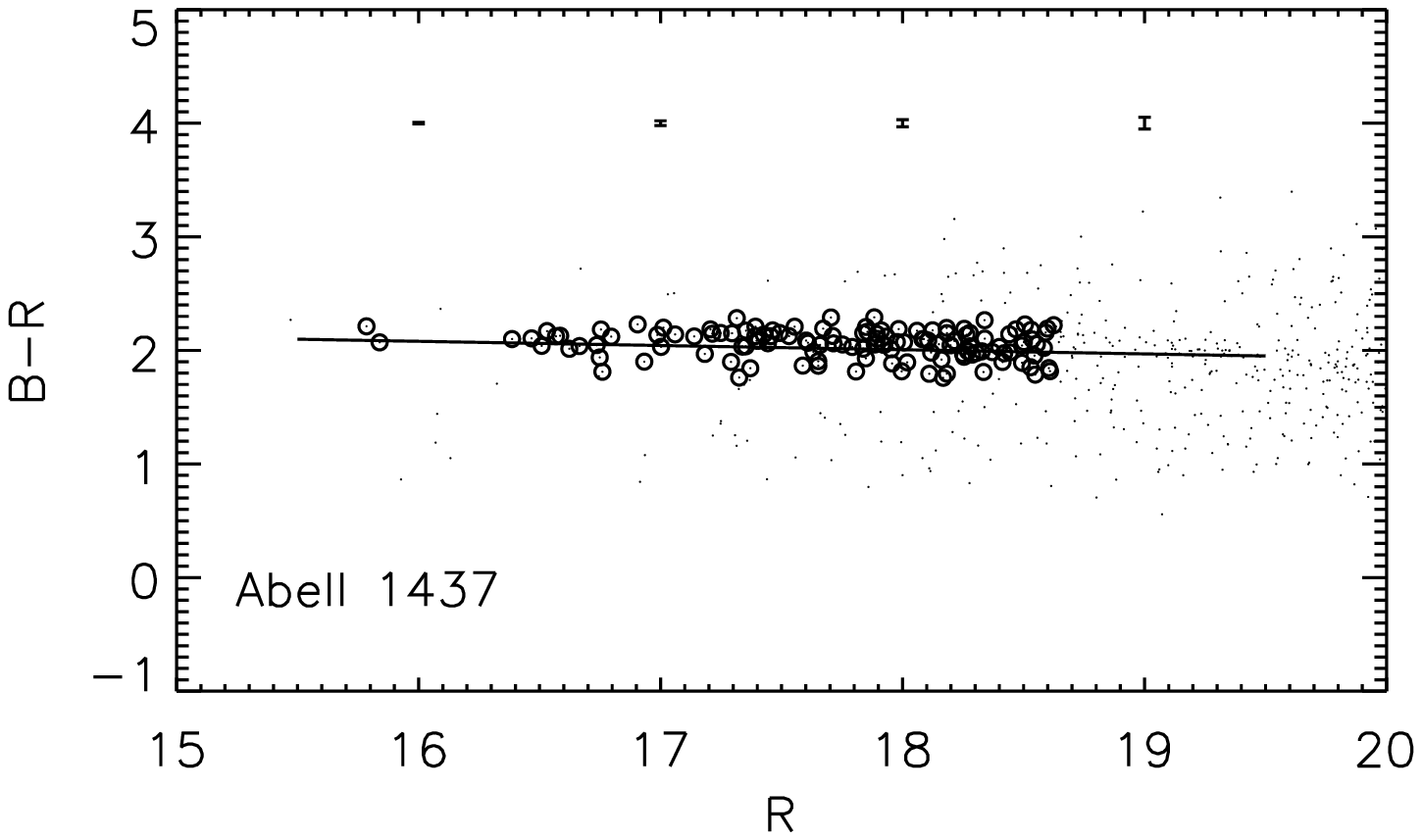}\includegraphics[width=0.4\textwidth]{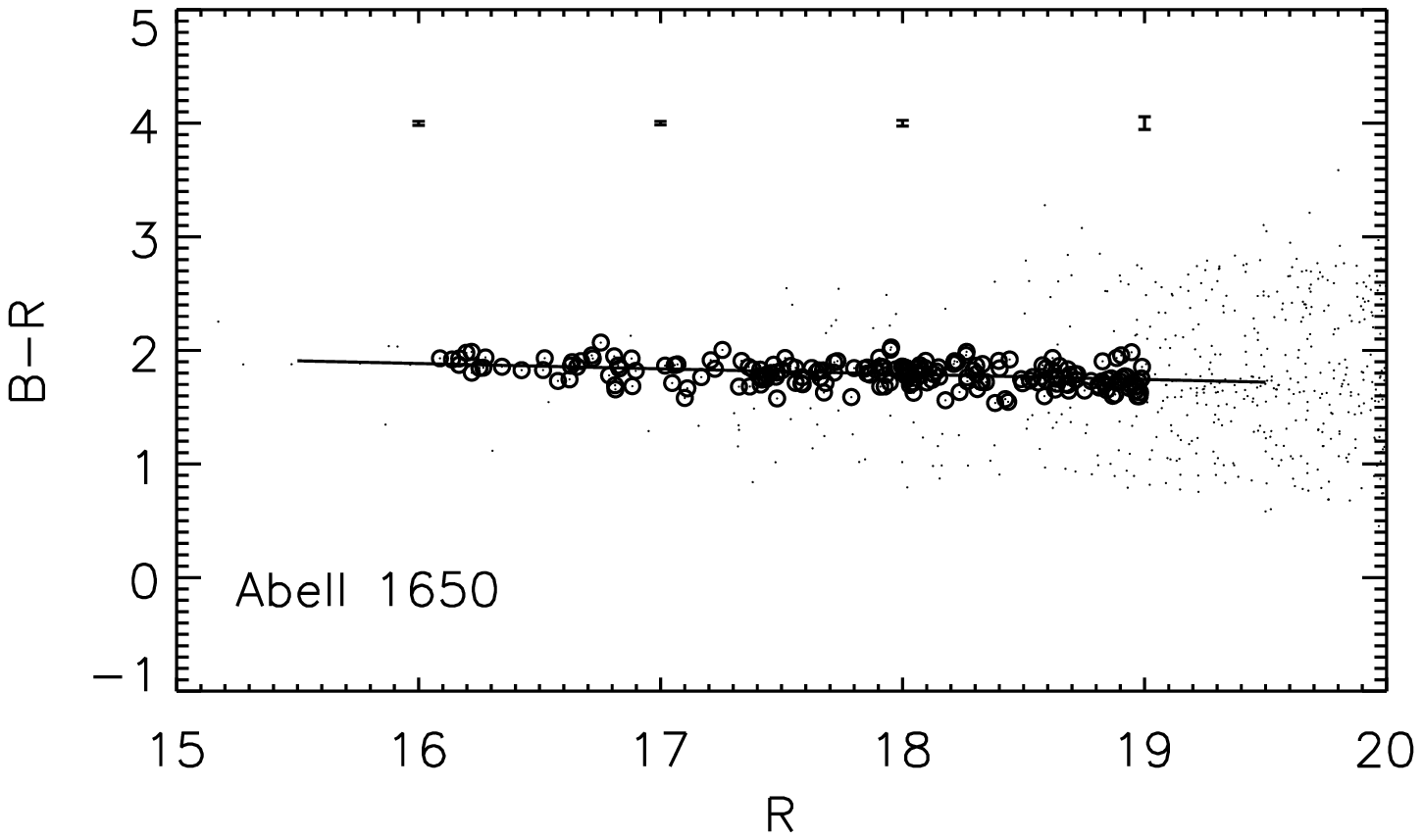}
\includegraphics[width=0.4\textwidth]{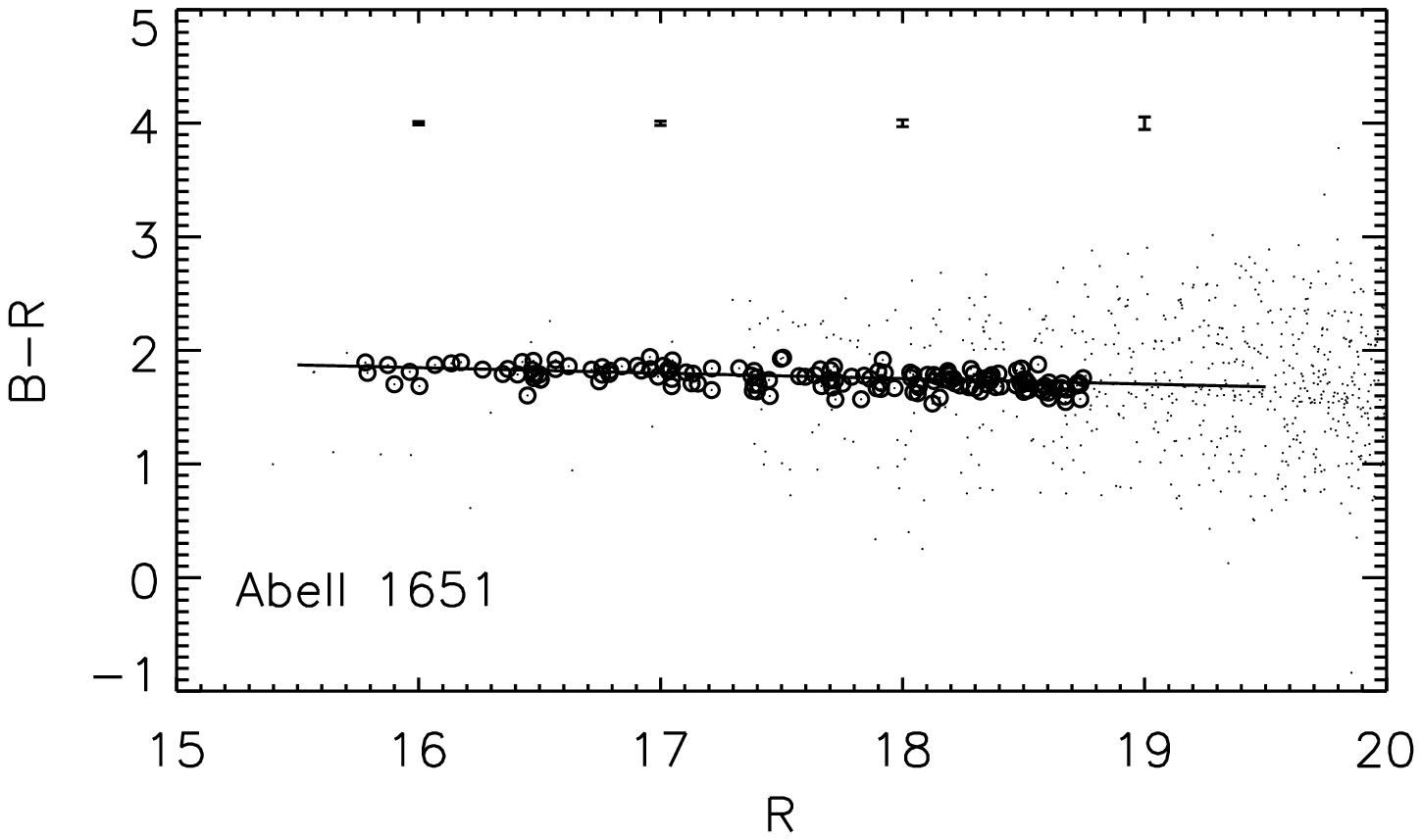}\includegraphics[width=0.4\textwidth]{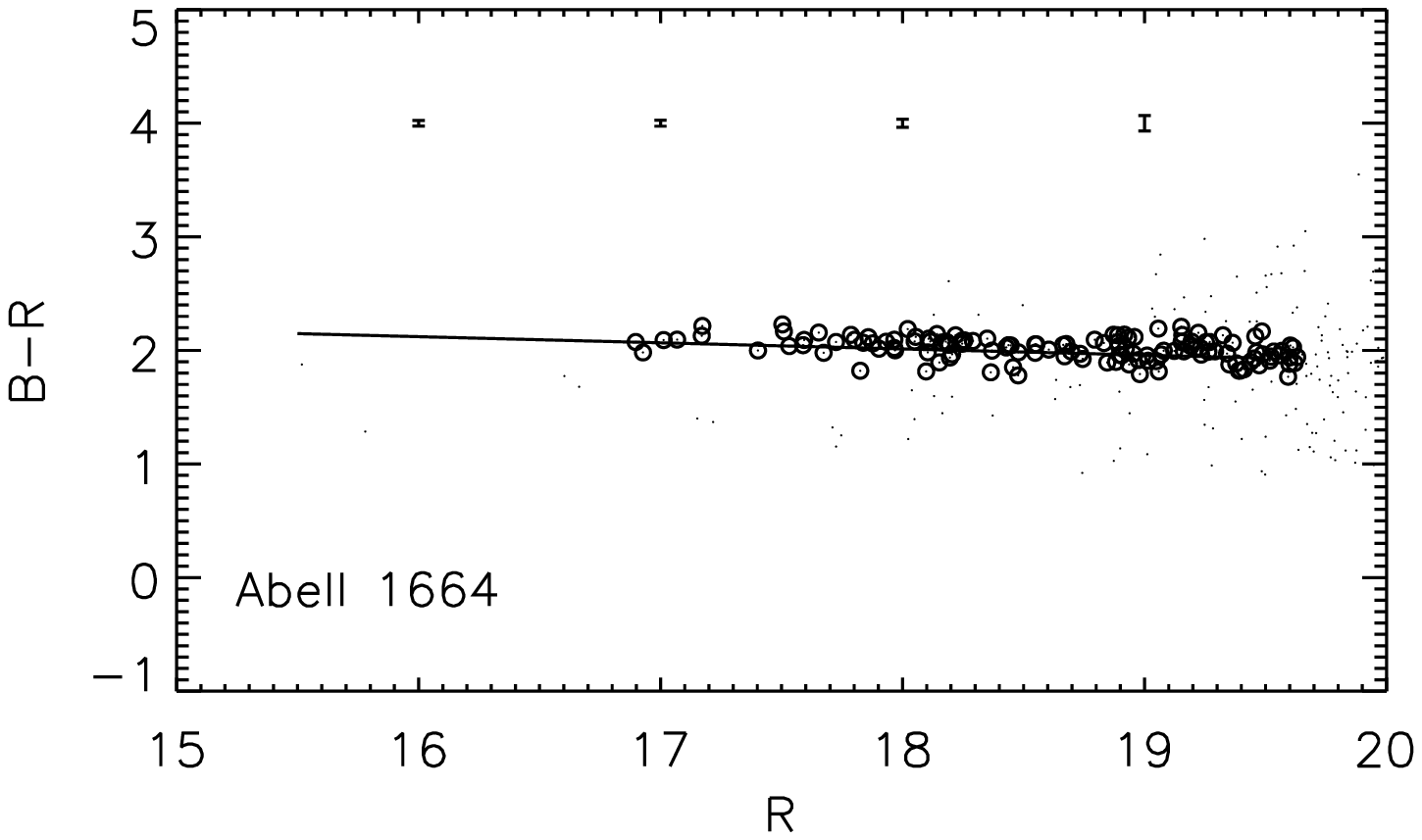}
\includegraphics[width=0.4\textwidth]{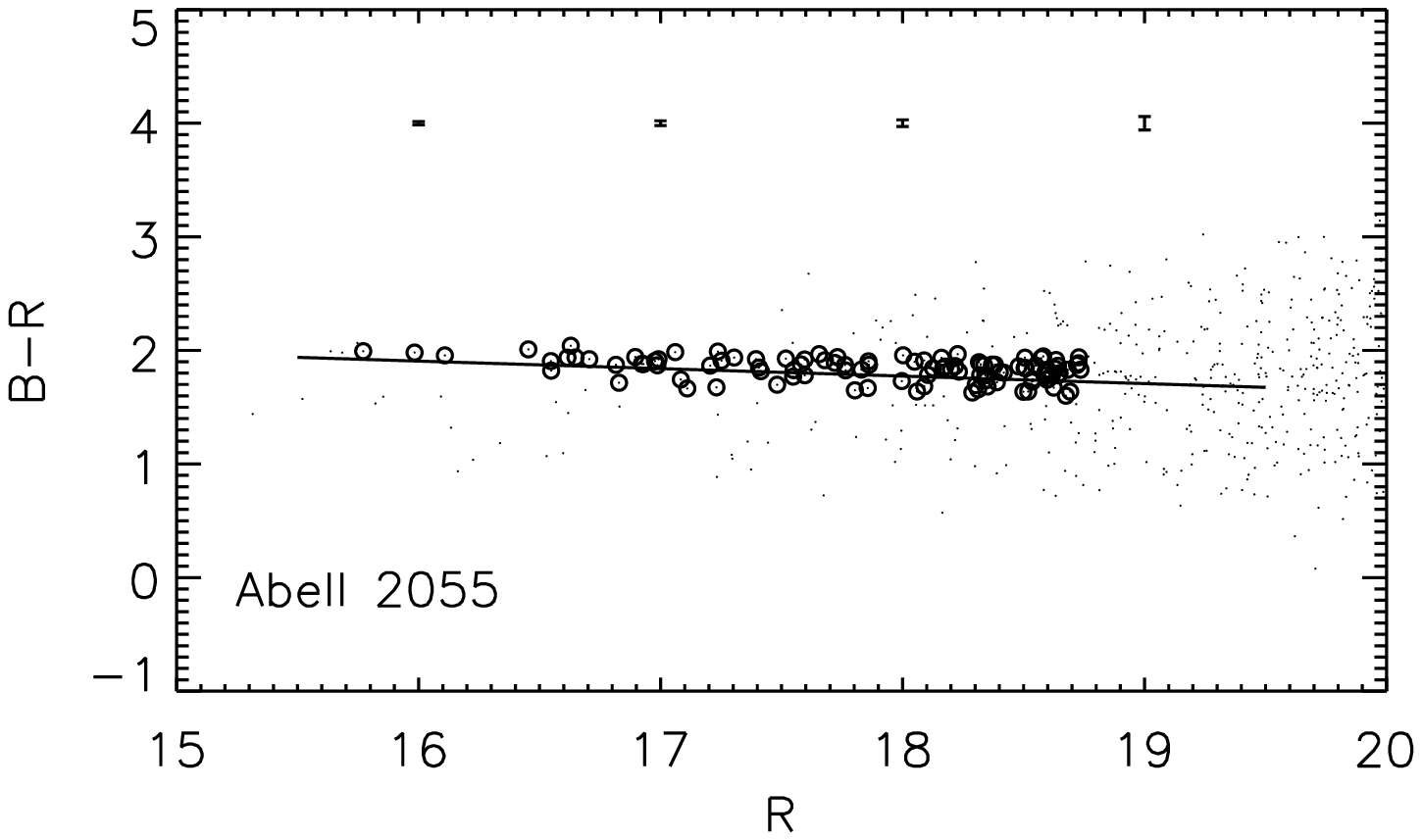}\includegraphics[width=0.4\textwidth]{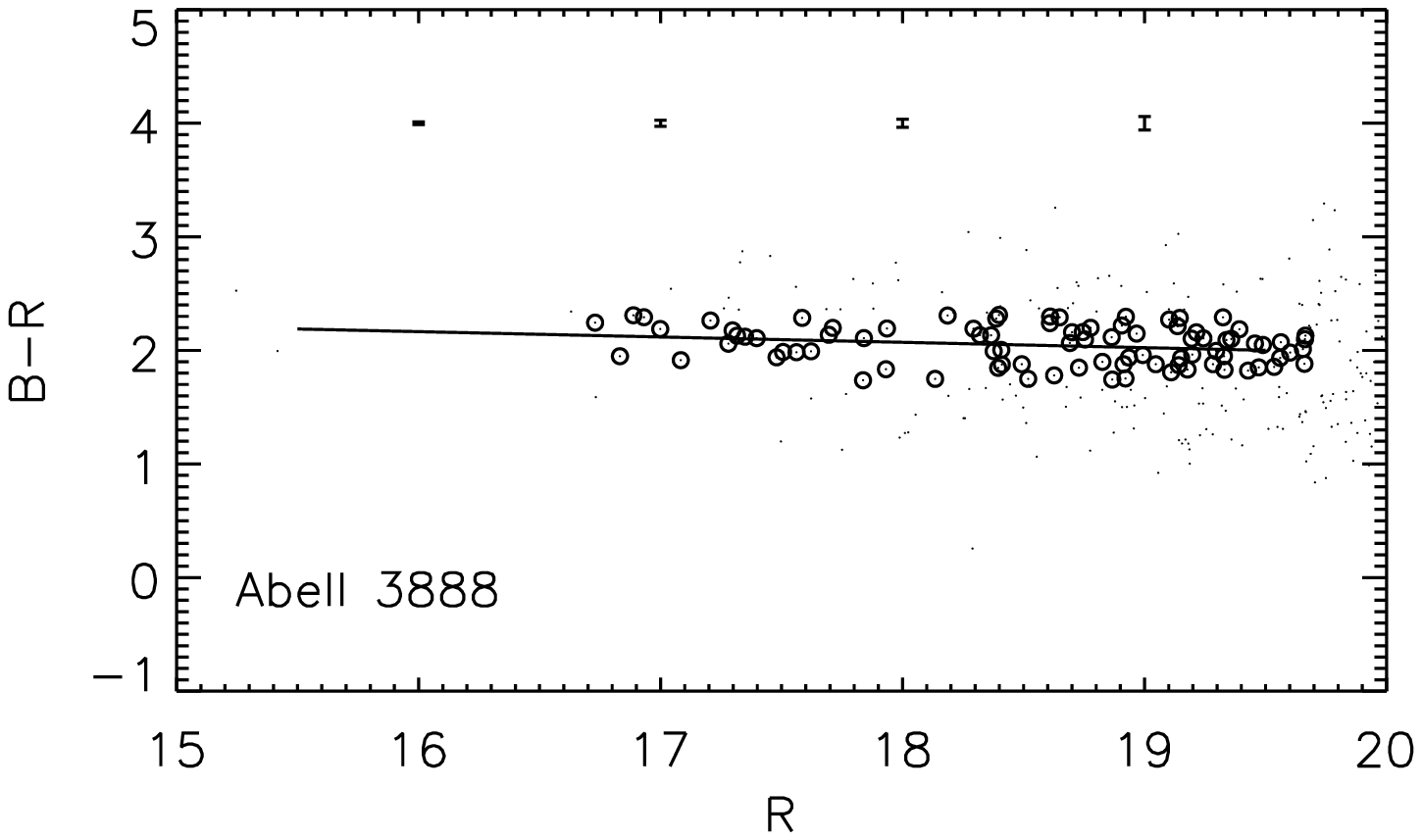}
\caption[]{The LARCS optical colour-magnitude diagrams with the red sequence fits overplotted. Representative error bars are shown for a range of magnitudes.}
\label{fig:larcscm}
\end{figure*}

\begin{figure*}
\centering
\includegraphics[width=0.4\textwidth]{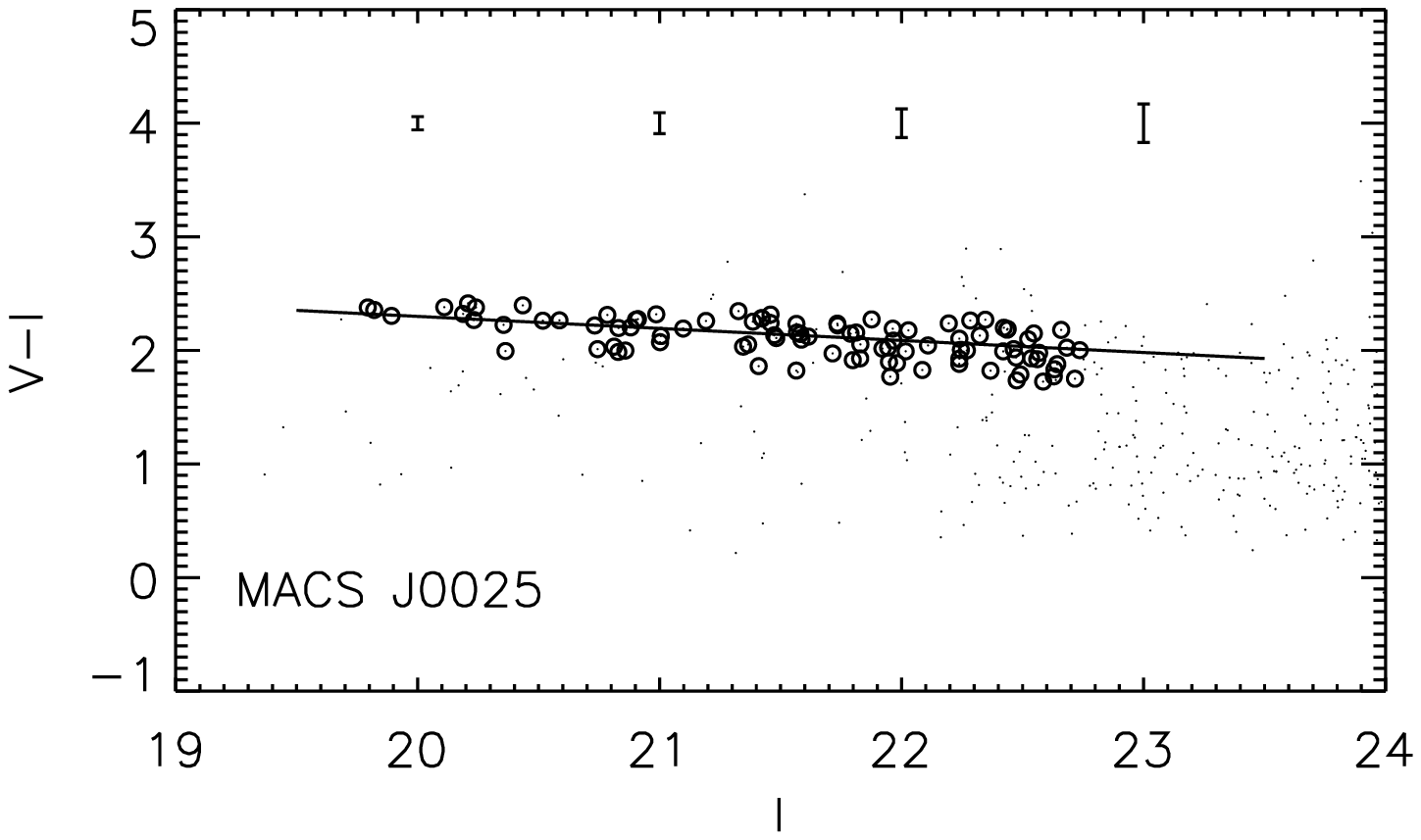}\includegraphics[width=0.4\textwidth]{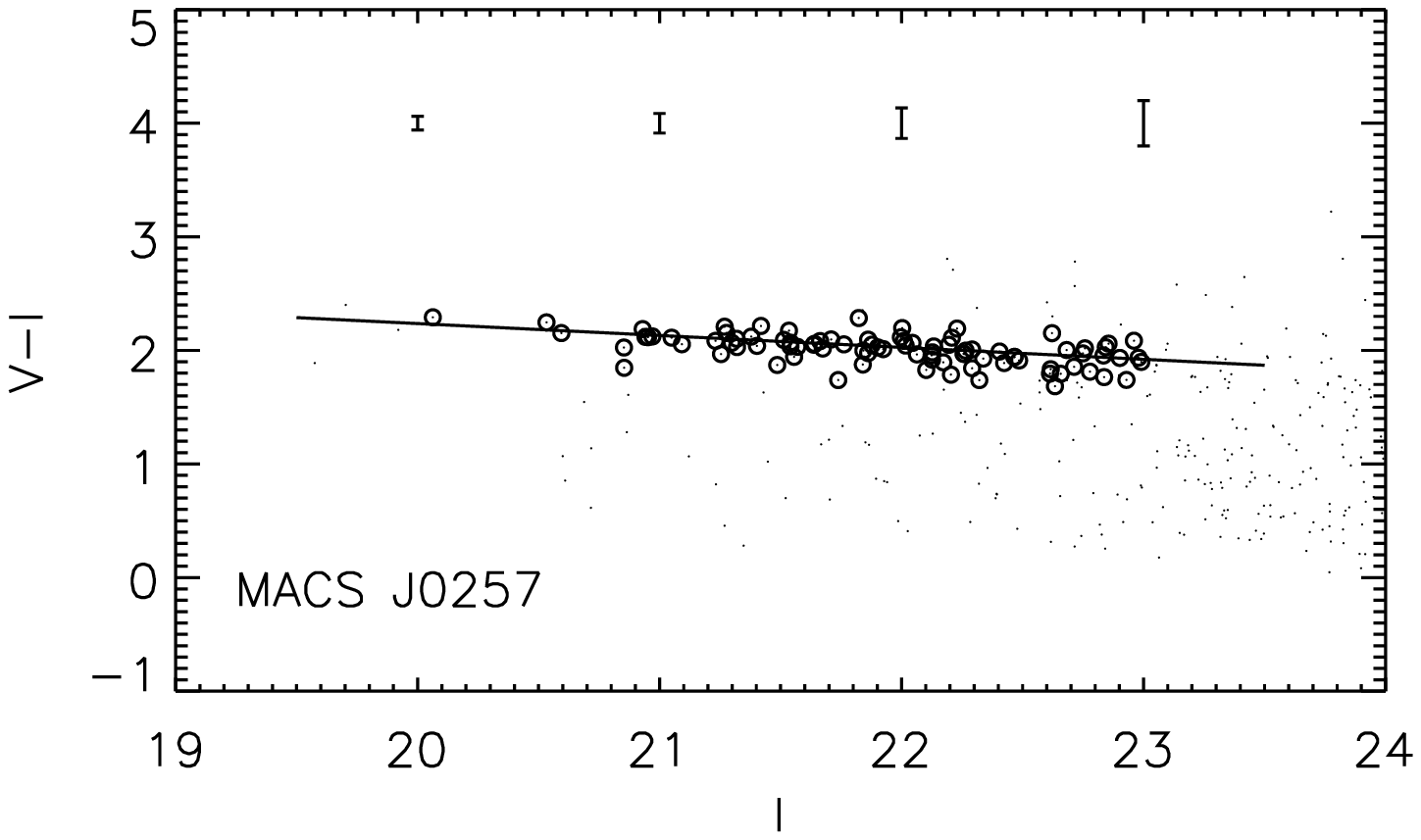}
\includegraphics[width=0.4\textwidth]{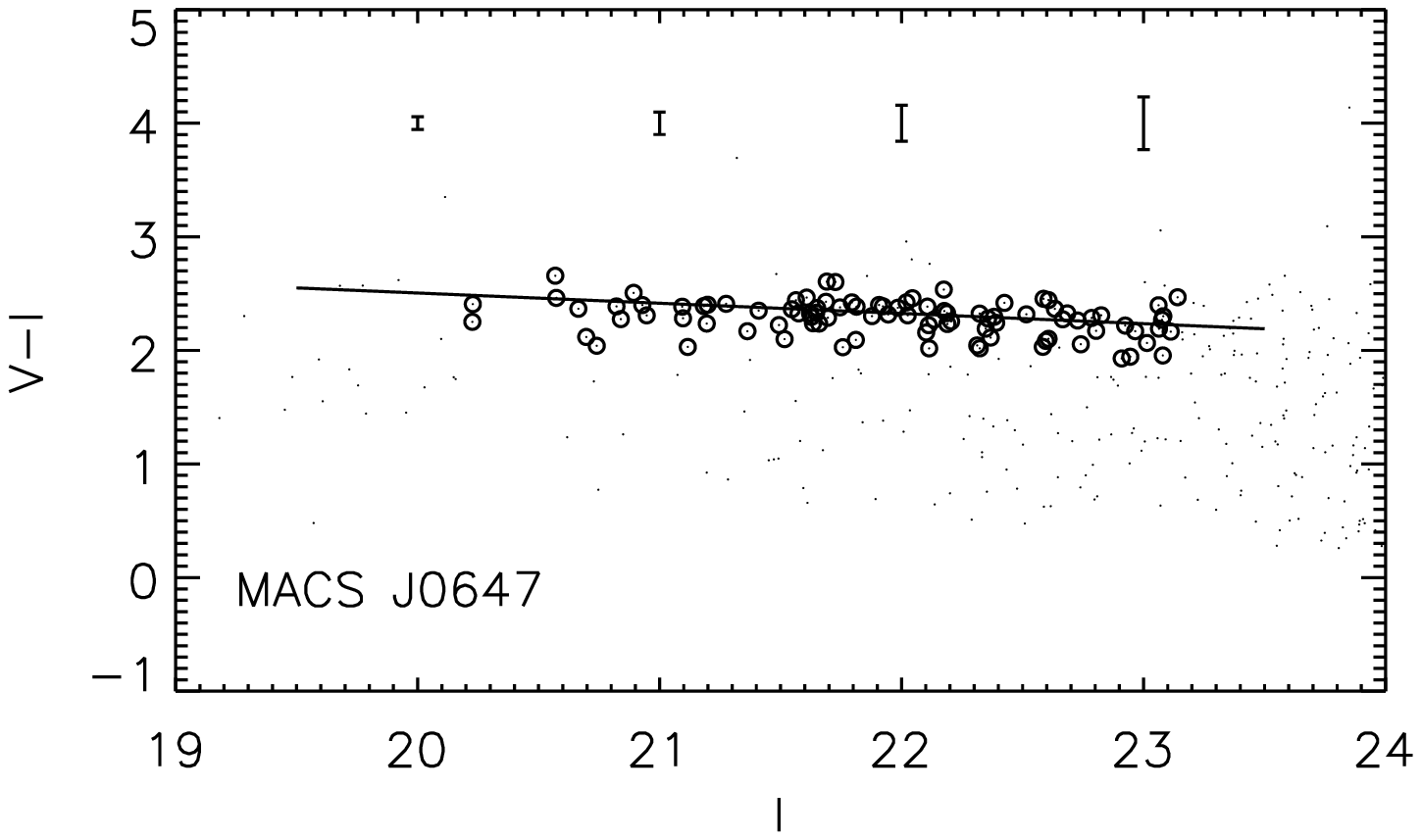}\includegraphics[width=0.4\textwidth]{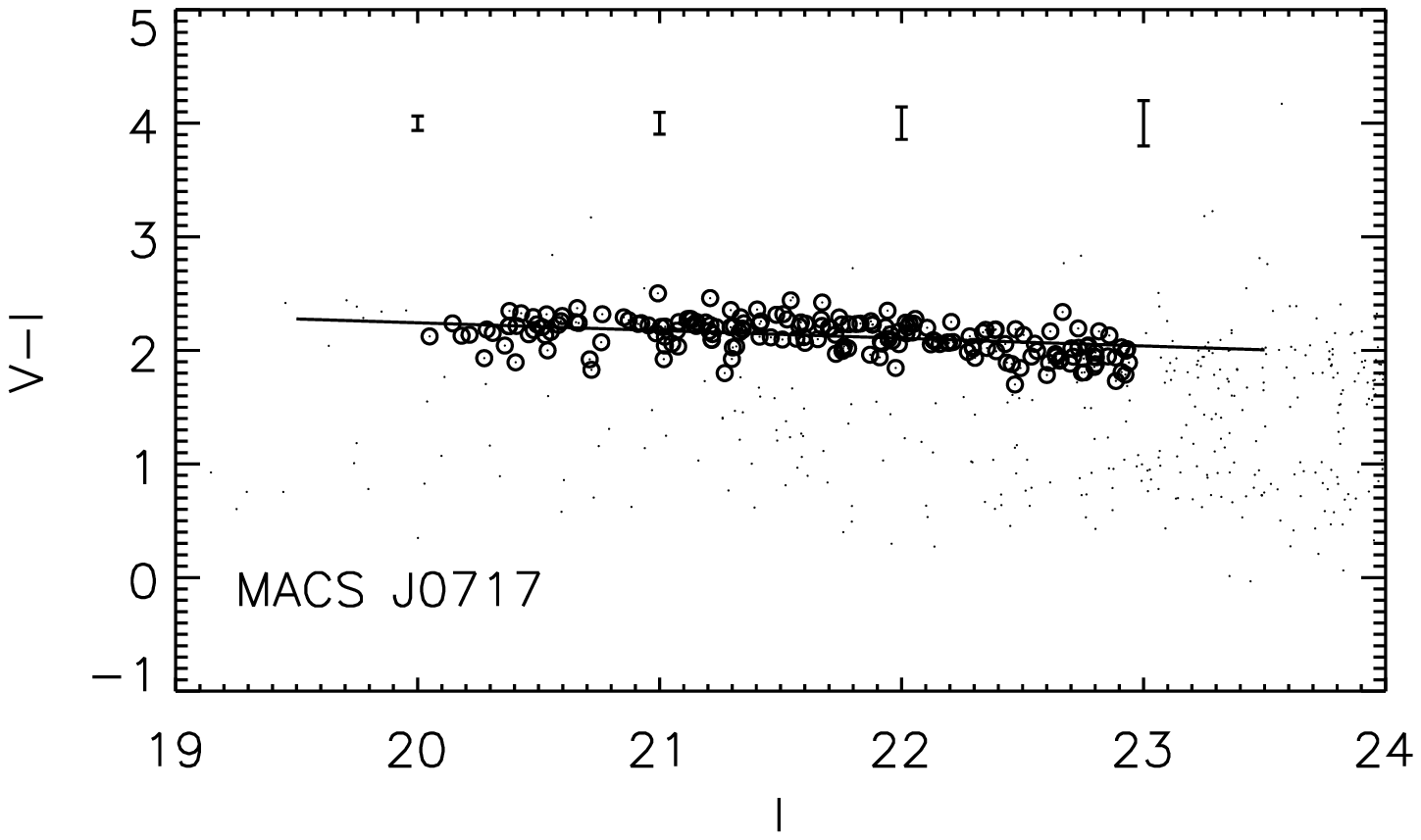}
\includegraphics[width=0.4\textwidth]{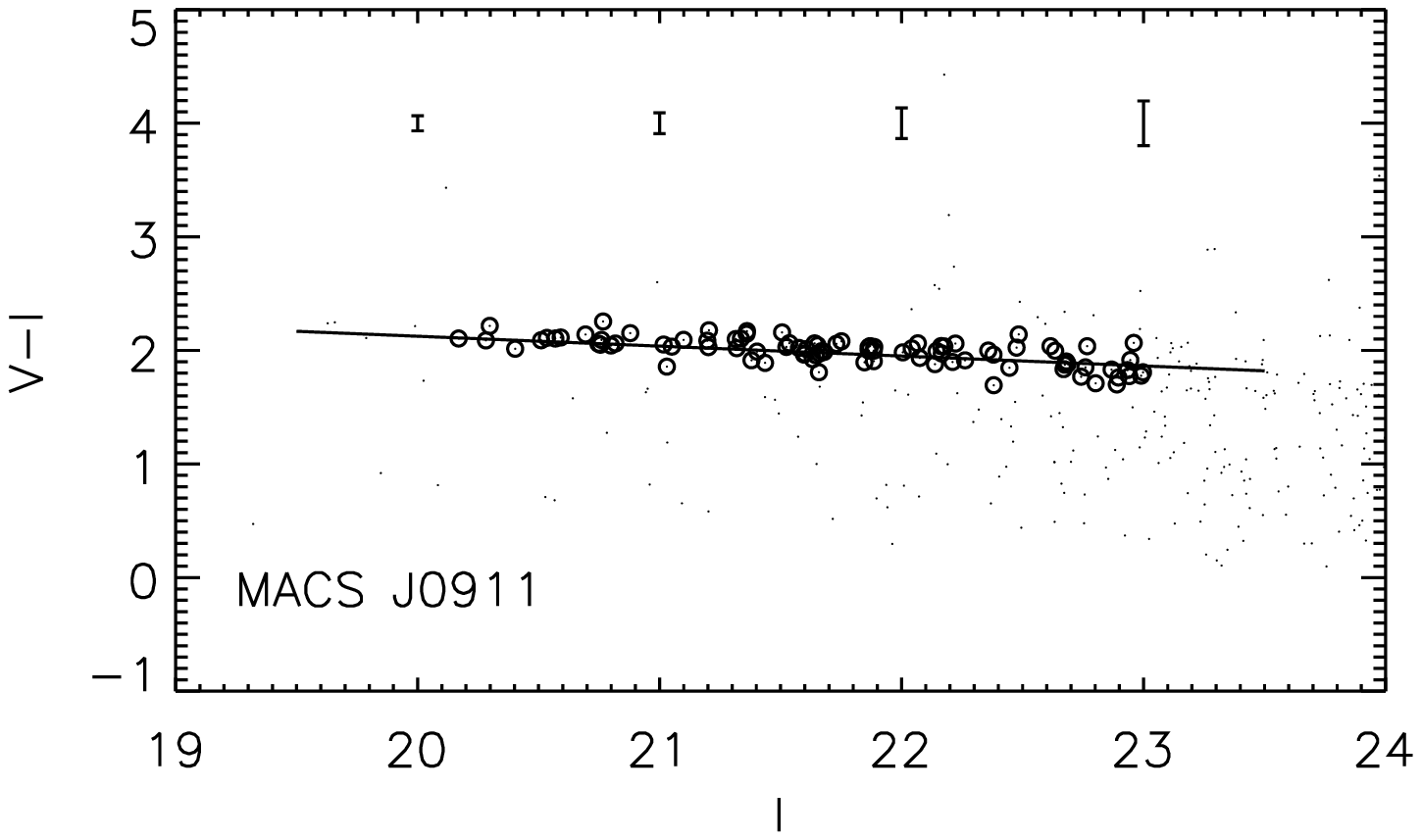}\includegraphics[width=0.4\textwidth]{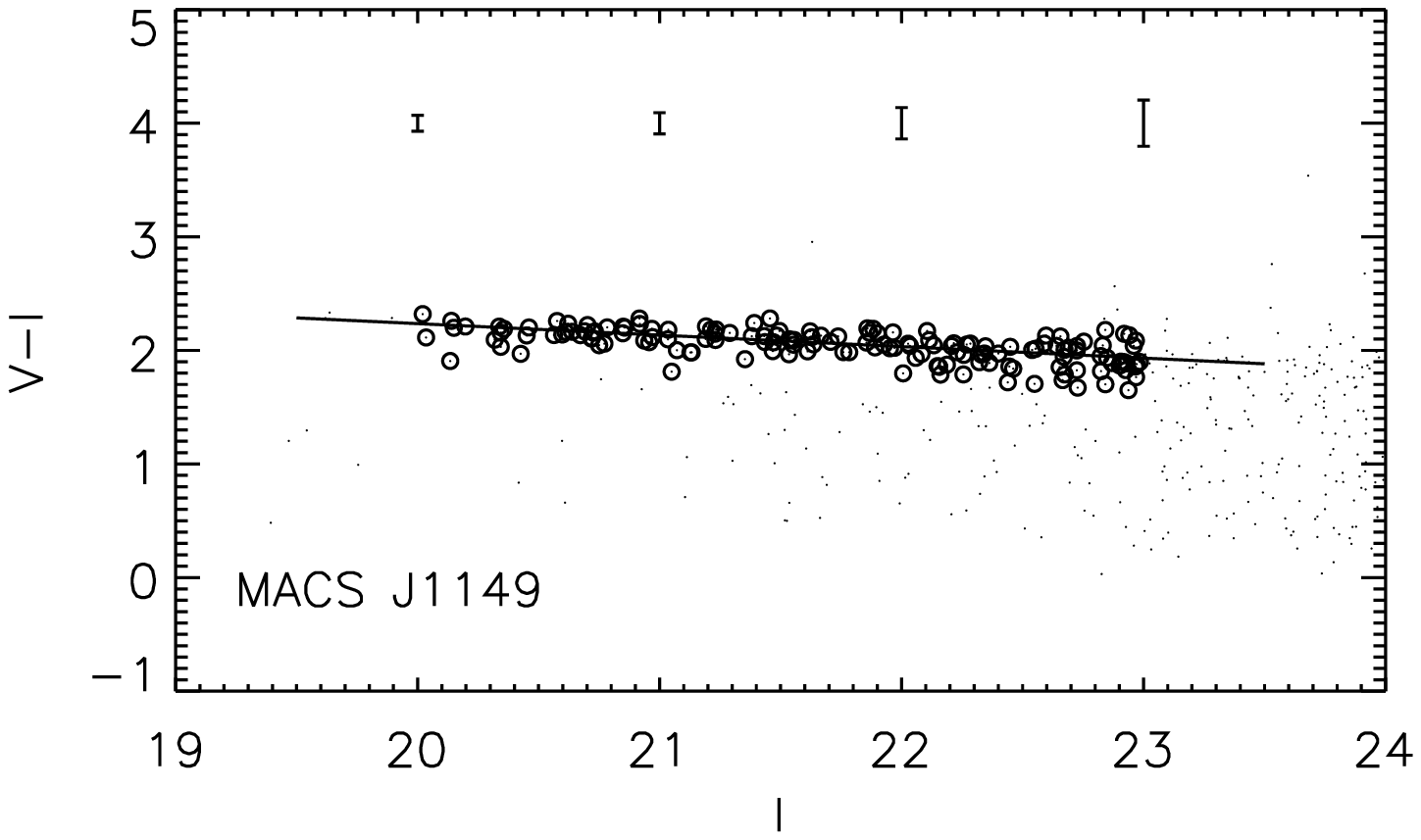}
\includegraphics[width=0.4\textwidth]{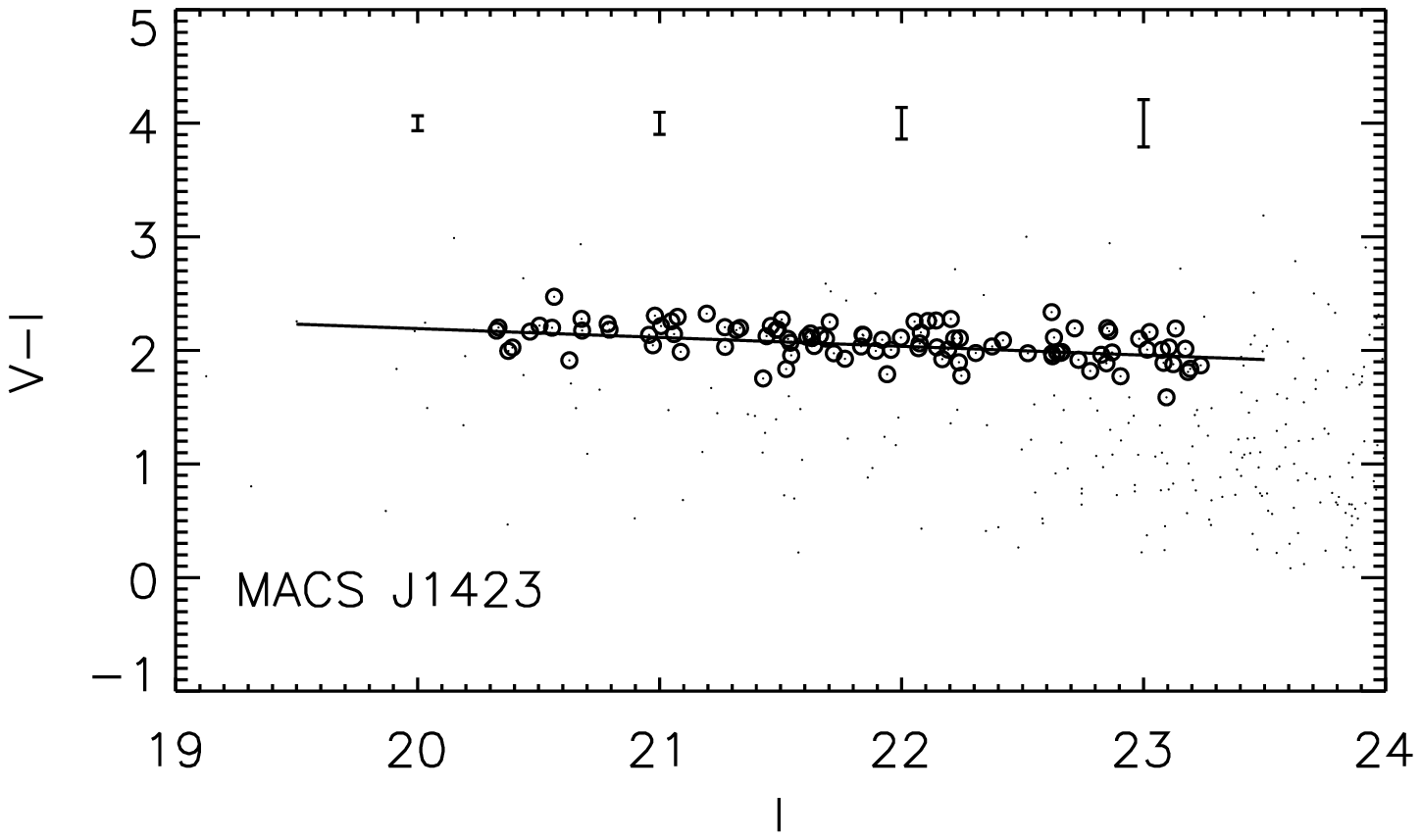}\includegraphics[width=0.4\textwidth]{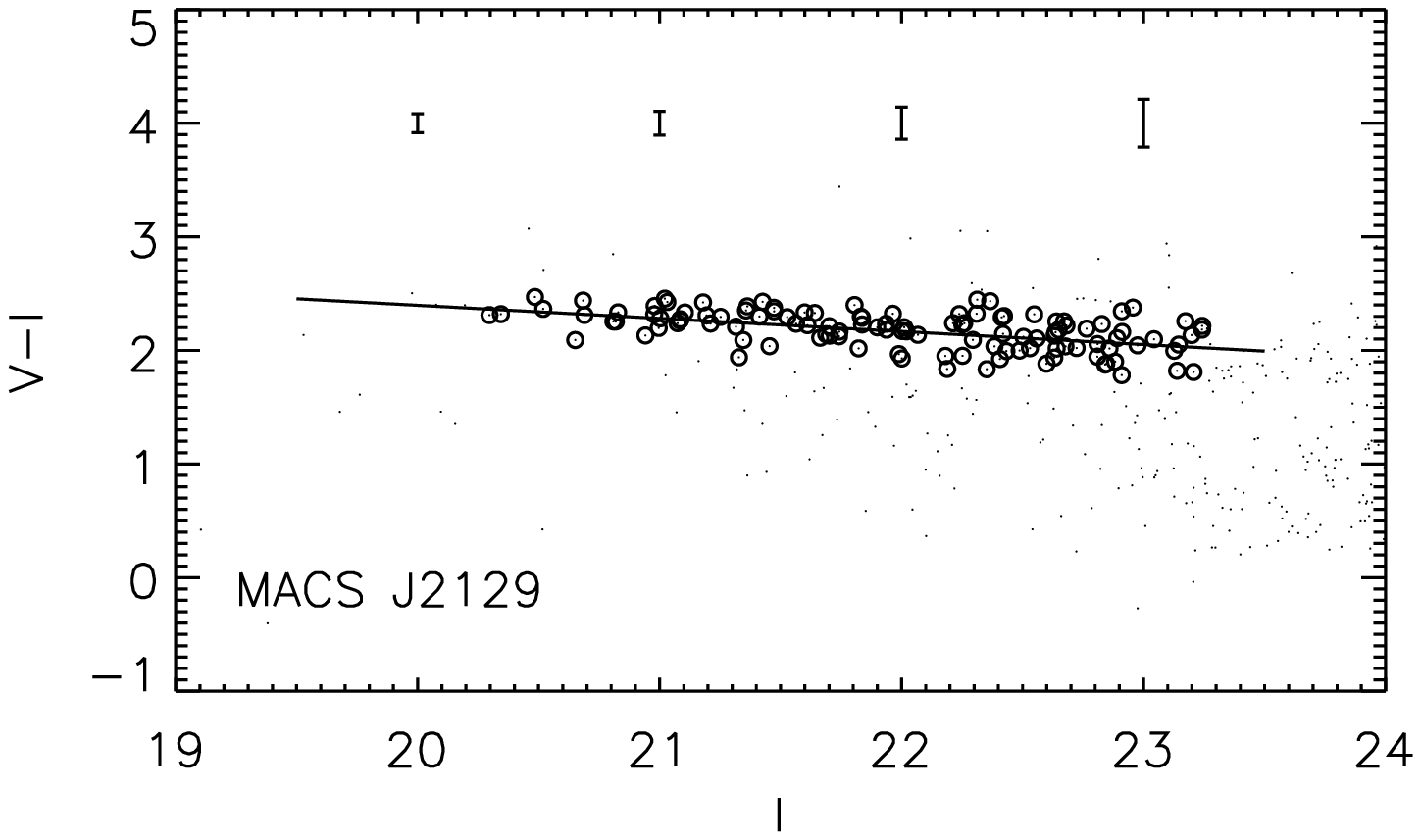}
\includegraphics[width=0.4\textwidth]{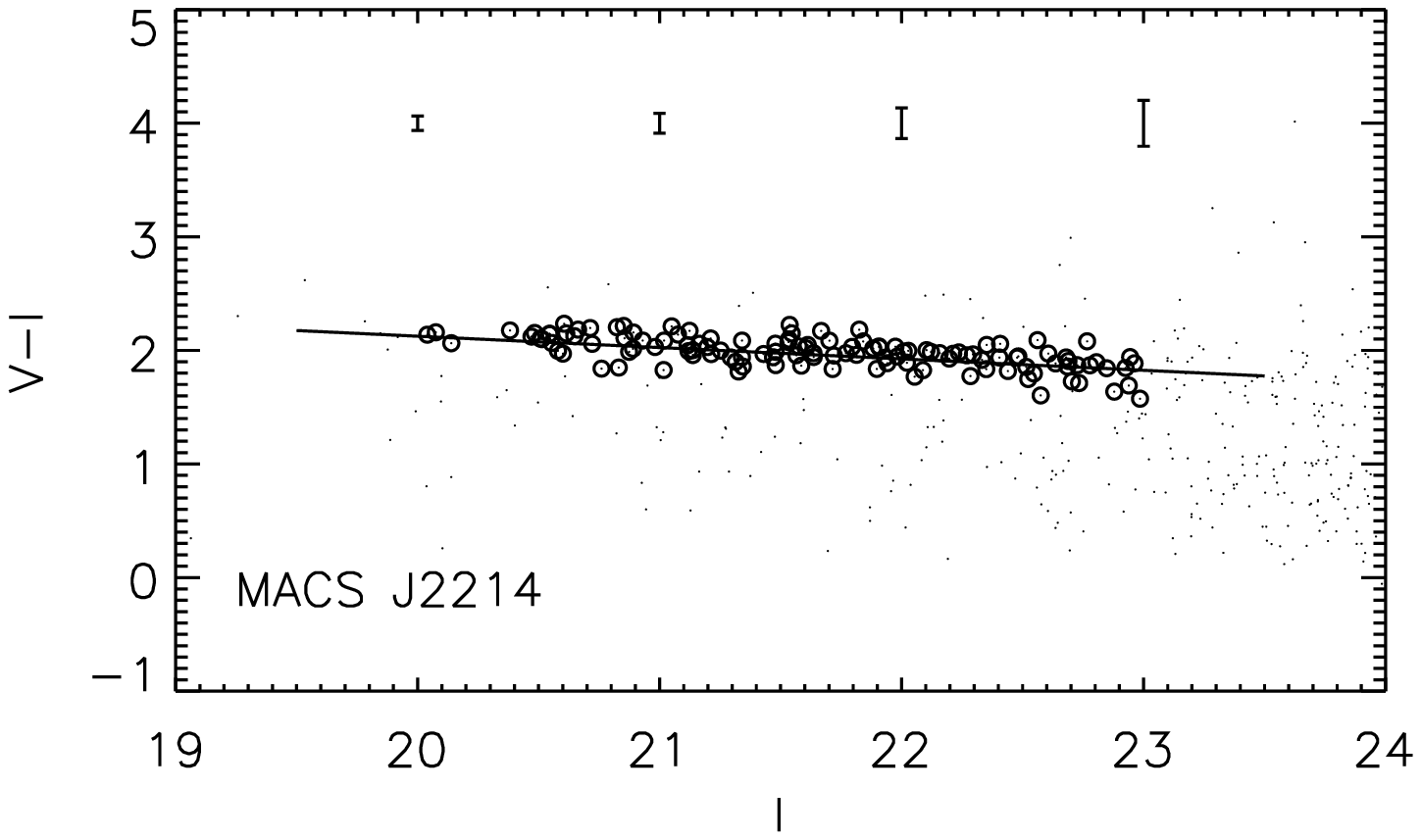}
\caption[]{The main MACS sample optical colour-magnitude diagrams with the red sequence fits overplotted. Representative error bars are shown for a range of magnitudes.}
\label{fig:macscm}
\end{figure*}

Figures \ref{fig:larcscm} and \ref{fig:macscm} are examples of the prominent red sequences we see in the rich clusters of our sample. To fit the slope of the red sequence we need to use a robust and consistent method. We therefore employ the same technique as described in \cite{Glad1998}, an iterated 3 $\sigma$ clipped fit. The fit is performed as follows: first we set a limiting magnitude for the red sequence which corresponds to the mean of the next 2 brightest galaxies down from the BCG magnitude+3. The reason for this is that in some cases there can be a large luminosity gap ($\sim$1 mag) between the BCG and the start of the red sequence so this ensures we measure the slope without this gap. The fit is performed over 3 magnitudes to incorporate the sub-L$^{\star}$ galaxy population which show the strongest evidence for red sequence build up \citep{Stott2007a}. We select the region of colour magnitude space containing the red sequence and estimate an initial fit from visual inspection. The residuals about this estimate are calculated and a Gaussian is fitted to the resulting colour distribution with the slope removed. The peak of this Gaussian corresponds to the red sequence. We then perform a fit to points that are within 3 sigma of this fit. This is a two parameter linear fit of the form $y = \kappa x + c$ where $\kappa$ is the slope of the red sequence. The process is iterated until it converges to a solution. We confirm the work of \cite{Glad1998} that this is a robust method for reasonable choices of limiting magnitude and initial fit.

It should be noted that our colour-magnitude diagrams are not field corrected as the rich clusters in our sample have well populated red sequences in contrast to the field. At the colours and magnitudes we consider this a very small contribution to the red sequence, typically 5\% contamination for the MACS clusters and 10\% for the LARCS clusters. We perform a statistical field correction test to a sub-sample of our low $z$ clusters and find that the slope of the sequence varies randomly by less than 1$\sigma$ than that obtained for the uncorrected sequence.  We therefore feel justified in not applying this correction. For details of the statistical field correction technique used see \cite{Stott2007a}.

\subsection{Model slope evolution}
\label{sec:modev}
We form a prediction of how the red sequence slope evolves, by analysing the semi-analytical model of \cite{Bower2006} based on the Millennium N-body simulation \citep{Springel2005}. We access these data online through Virgo Millennium Database. The \cite{Bower2006} model accounts for the recent observations that the stellar mass in bright galaxies was in place at high redshift by including feedback from active galactic nuclei. It also includes a prescription for the build of the red sequence via `strangulation'. This describes a process whereby galaxies falling into a cluster are stripped of their hot gas reservoir upon interaction with the intracluster medium quenching star formation \citep{larson1980}. 

We define model clusters as dark matter haloes above a threshold mass which contain bound sub-haloes (galaxies) with observational properties assigned to them by semi-analytic modelling. To allow comparison between models and observation we assume that the model cluster members are analogous to the red sequences of our observed clusters. We need to select clusters from \cite{Bower2006} which are comparable in mass to those we observe. By using the relation between cluster mass and X-ray luminosity \citep{popesso2005} we include only the simulated haloes with M$_{200}>3.4\times10^{14}$M$_{\odot}$  which corresponds to L$_{X}>10^{44}$erg s$^{-1}$. To select only the most massive systems for comparison with our observations we limit our red sequence fit to a stack of the top five ranked clusters by mass in each redshift bin. The model slope is robust to our choice of lower mass limit as we see no significant change in our slope values when using all haloes with masses greater than mean halo mass in the required redshift interval.

We model the red sequence slope evolution by creating stacked colour magnitude diagrams from the \cite{Bower2006} model output at a distinct set of redshift intervals between $z=0$ and $z=1$. For the creation of these colour-magnitude diagrams we ensure we only study the passive red sequence galaxies by selecting galaxies with no current star formation (L(H$\alpha)=$0). The models provide no spatial information but we assume that the synthetic colours we calculate from the total magnitudes can be compared with the observed aperture magnitudes allowing for a normalisation between the model and observed slope evolution at low redshift.  

We calculate the model for the observed frame slope evolution and its errors by fitting to the stacked synthetic red sequence slopes at each redshift interval with the method described in \S\ref{sec:fit}. As with the observations, the model slope is shown to steepen with redshift. 

\subsection{Observed slope evolution}
\label{sec:sev}
\begin{figure*}
\centering
\includegraphics[width=0.6\textwidth]{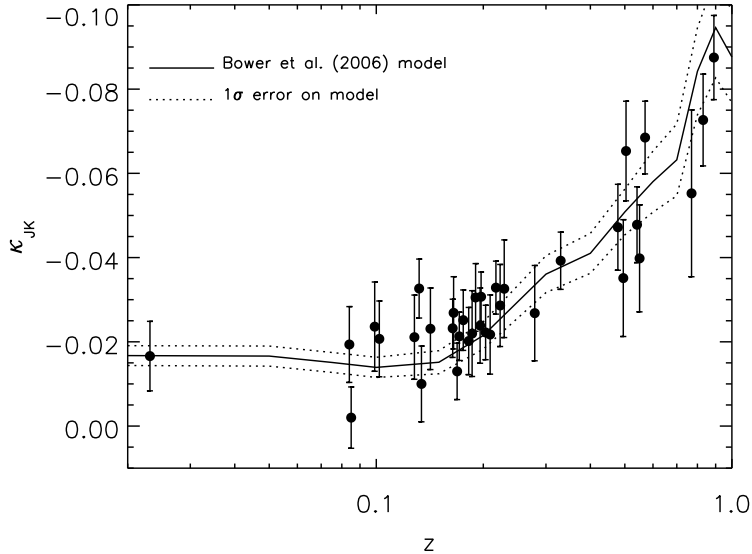}
\caption[The evolution of the red sequence slope ($\kappa_{JK}$) for our near-infrared sample.]{The evolution of the red sequence slope ($\kappa_{JK}$) for our near-infrared sample. A model calculated from \cite{Bower2006} is included for comparison with theory}
\label{fig:evl}
\end{figure*}

Fig. \ref{fig:evl} displays the observed red sequence slope evolution for our near infrared sample. The slope ($\kappa_{JK}=\delta(J - K)/\delta K$) is shown to steepen with redshift. This steepening will have contributions from both $K$ correction and perhaps an age or a metallicity evolution. The contribution from $K$ correction is due to the observed J and K bands sampling increasingly bluer rest wavebands at higher redshift which affects galaxies differentially along the red sequence depending on their spectral energy distributions (\citealt{Glad1998}; \citealt{LARCS2001}). 

When the model described in \S\ref{sec:modev} is plotted with our complete near-infrared dataset in Fig.\ref{fig:evl} we find good quantitative agreement between the two with the data having an rms scatter about the model of 0.009 which is comparable to the calculated error on the model (mean 1$\sigma$ error is  0.006).  

\begin{figure*}
\centering
\includegraphics[width=0.6\textwidth]{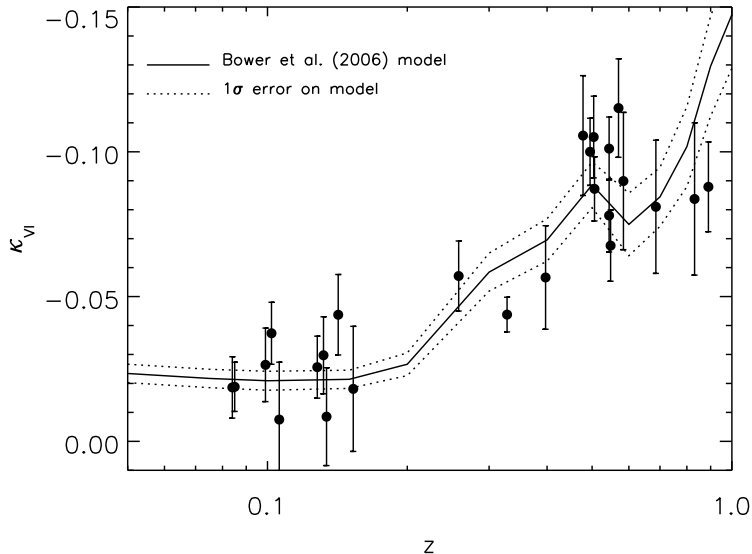}
\caption[The evolution of the red sequence slope ($\kappa_{VI}$) for our optical sample.]{The evolution of the red sequence slope ($\kappa_{VI}$) for our optical sample. The model again comes from analysis of \cite{Bower2006}}
\label{fig:evr}
\end{figure*}

\begin{figure*}
\centering
\includegraphics[width=0.6\textwidth]{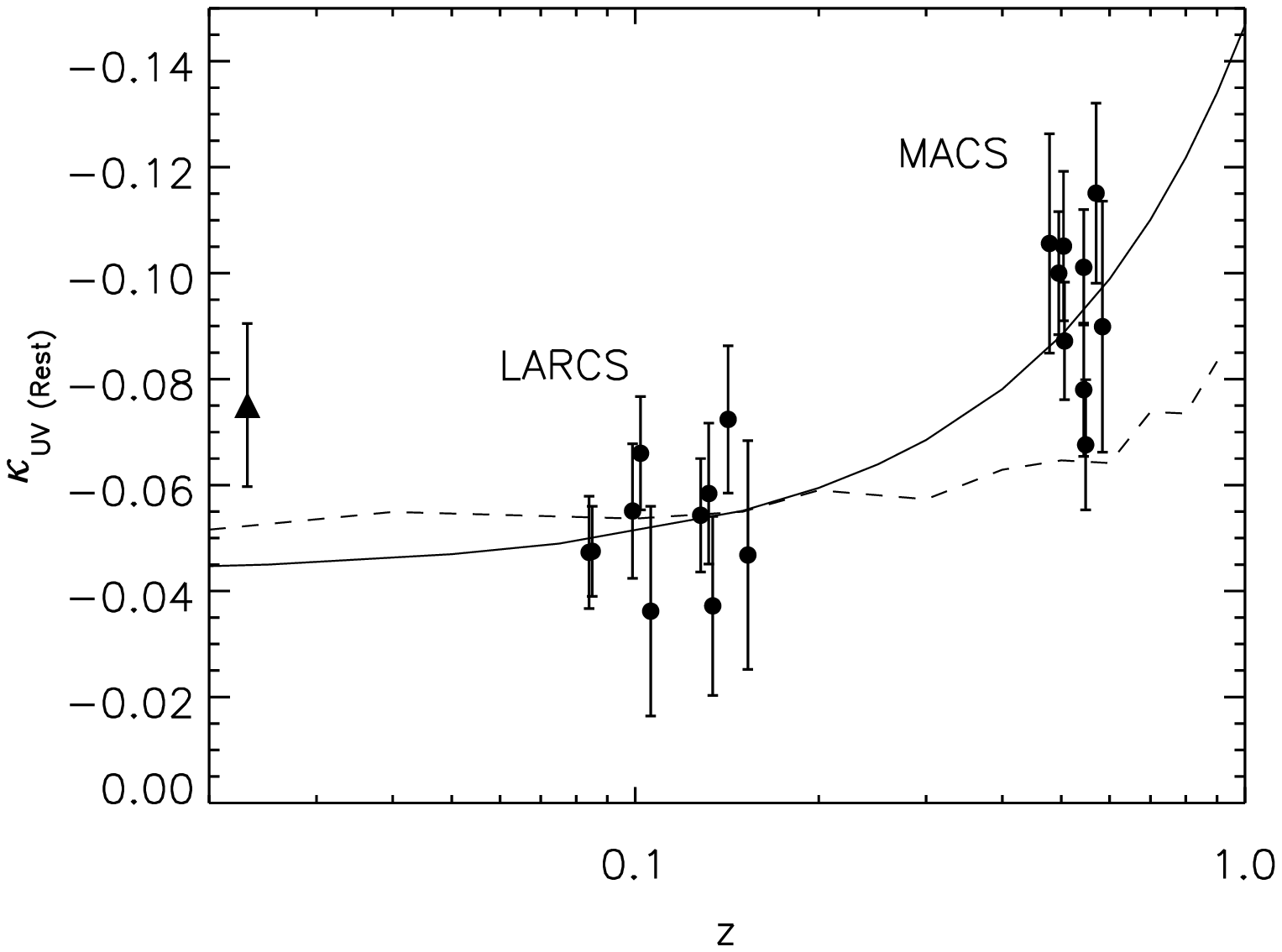}
\caption[]{The evolution of the rest frame red sequence slope ($\kappa_{UV}$) for our optical sample. The triangle is the data point for the Coma Cluster. The solid line is a fit to the MACS and LARCS data of the form  $(1+z)^{\beta}$ where $\beta=1.77\pm0.25$. The dashed line is the rest frame slope evolution calculated from the semi-analytic model of \cite{Bower2006}.}
\label{fig:restev}
\end{figure*}

We now investigate the slope ($\kappa_{VI}=\delta(V - I)/\delta I$) evolution for our optical observations. As above, the simulated slope evolution plotted is calculated from semi-analytical model of \cite{Bower2006}. There is an additional complication as our optical data are sourced from two different filter sets so we have to account for the observed difference in red sequence slope between them. To achieve this we normalise all data to the $\delta(V - I)/\delta I$ filters by correcting for the difference between the $\delta(V - I)/\delta I$ model and the model for the $\delta(B - R)/\delta R$ filter combination (as in \citealt{Glad1998}). The resultant data-points and model are plotted in Fig. \ref{fig:evr}. The slope is shown to increase with redshift as above. The rms scatter about the model is 0.017 (compared to the mean 1 $\sigma$ error in the model of 0.008) so as with the near-infrared observations we find good agreement between the model and the data. 

\subsubsection{Rest frame slope evolution}
\label{sec:restf}
To investigate the intrinsic evolution of the red sequence slope we need to study clusters observed with matched rest frame photometry. This is to quantify the proposed contribution from the build up of the red sequence without the additional effect of $K$ correction. We can observe this intrinsic slope evolution as our two main sets optical observations, the $z\sim$0.1 LARCS clusters in $(B - R)$ colour and the $z\sim$0.5 MACS clusters in $(V - I)$ colour, both correspond to the rest frame $(U - V)$ colour at their respective epochs. $(U - V)$ colour straddles the 4000\AA\, break and therefore is a good discriminant between the red sequence and star forming cluster members/foreground galaxies. We confirm that the filters are well matched to rest frame $(U - V)$ as we find a colour term of 1.0 $\pm$0.05 between the $B - R$ at $z=$0.11 and $V - I$ at $z=$0.53 using the technique described in \cite{blake2006}.

The evolution of this intrinsic slope ($\kappa_{UV}=\delta(U - V)/\delta V$) is plotted in Fig. \ref{fig:restev}.  We include an additional low redshift data-point for the Coma cluster calculated from the Sloan Digital Sky Survey (SDSS, \citealt{SDSS}) $u$ and $g$ filter photometry. In this figure we can see that the intrinsic optical slope, $\kappa_{UV}$, does evolve with redshift as the intermediate $z$ MACS clusters have a steeper red sequence than their low $z$ LARCS counterparts. The weighted mean values for the LARCS and MACS $\kappa_{UV}$ are -0.053$\pm$0.004 (s.e.m) and -0.092$\pm$0.004 (s.e.m) respectively, a difference of 6.5 $\sigma$. We can therefore say that there is a real contribution to the slope evolution from factors other than $K$ correction. The fit to the data in Fig. \ref{fig:restev} is of the form $(1+z)^{\beta}$ where $\beta=1.77\pm0.25$. We note that the Coma Cluster has a steeper slope than the rest of our low redshift sample, although only a $\sim$2$\sigma$ discrepancy from the fit, which we may expect as it is found to have lower than average dwarf-to-giant ratio along its red sequence suggesting it is still undergoing faint end and therefore slope evolution \citep{Stott2007a}. Including the Coma Cluster does not have a significant affect on the fit, $\beta=1.67\pm0.26$. 

The dashed line plotted on Fig. \ref{fig:restev} is the rest frame $\kappa_{UV}=\delta(U - V)/\delta V$ slope calculated from the semi-analytic model of \cite{Bower2006}, normalised to the mean value of the LARCS sample slope, which shows only a mild evolution with redshift and is therefore unable to replicate the rest frame slope change observed. From this we can conclude that the major contribution to the agreement between the observed slope evolution in $\S\ref{sec:sev}$ and that derived from the semi-analytic modelling shown in Figs \ref{fig:evl} \& \ref{fig:evr} was the $K$ correction differential between high and low mass galaxies along the synthetic red sequence and that a significant, intrinsic, slope evolution is not predicted by the models.  

\subsection{Evolution with other observables}
\label{sec:xev}
\begin{figure}
\centering
\includegraphics[width=0.5\textwidth]{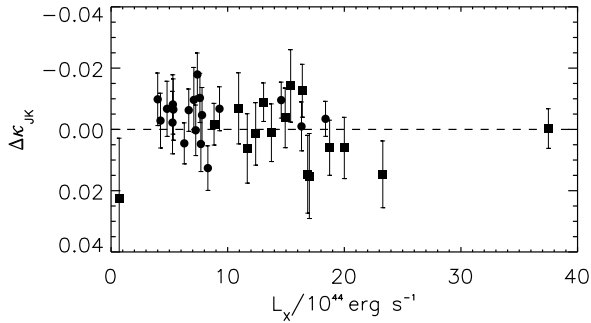}
\caption[]{The residuals about the models in Fig. \ref{fig:evl} plotted against X-ray luminosity. Filled circles and squares represent galaxy clusters with redshifts, $z<0.2$ and $z>0.2$ respectively, to demonstrate there is no redshift dependence. The correlation coefficient, r, for this plot is 0.17.}
\label{fig:reslxir}
\end{figure}

\begin{figure}
\centering
\includegraphics[width=0.5\textwidth]{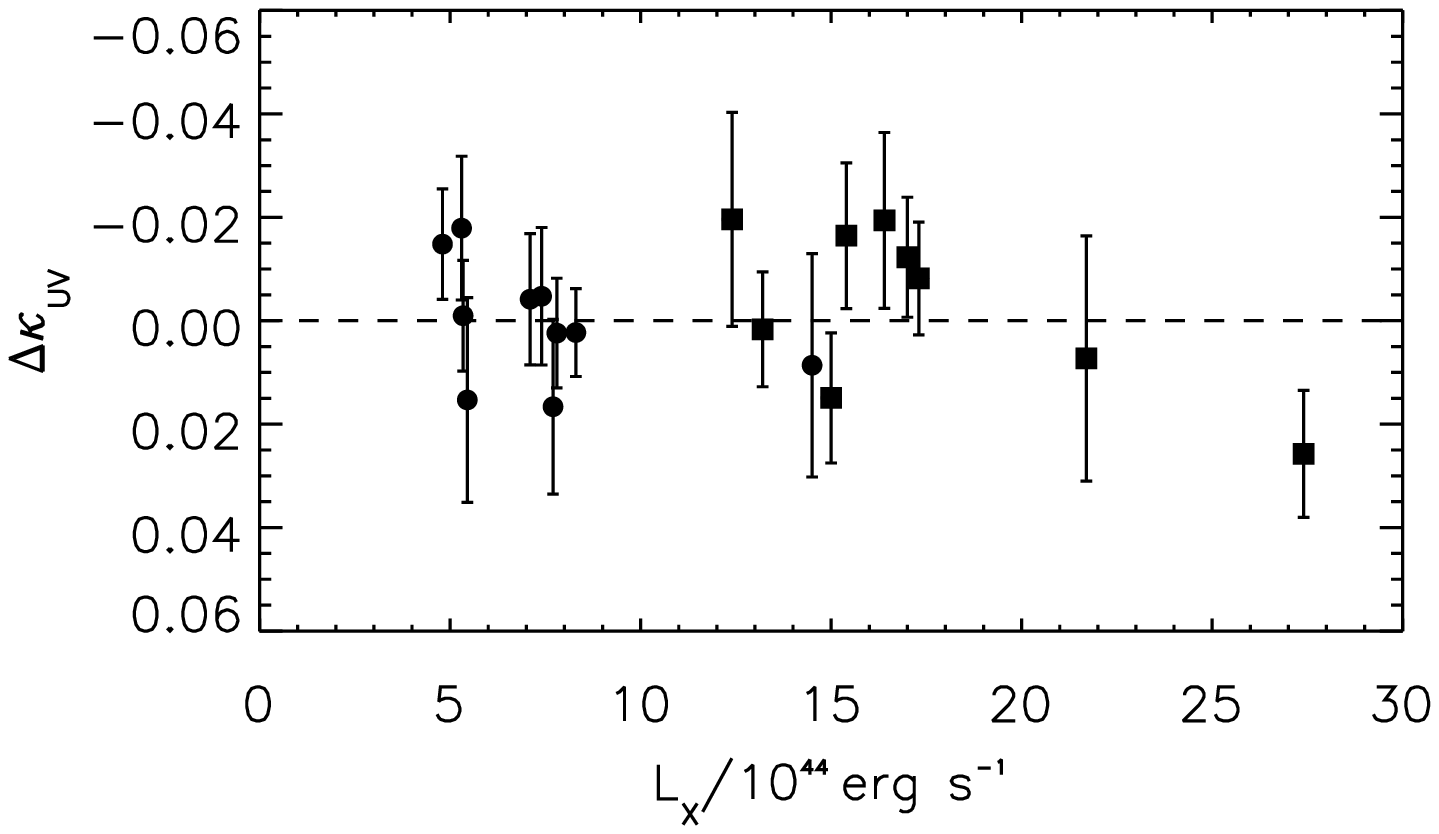}
\caption[]{The residuals about the models in Fig. \ref{fig:restev} plotted against X-ray luminosity. Filled circles and squares represent galaxy clusters with redshifts, $z<0.2$ and $z>0.2$ respectively, to demonstrate there is no redshift dependence. The correlation coefficient, r, for this plot is 0.24.}
\label{fig:reslxuv}
\end{figure}

\begin{figure}
\centering
\includegraphics[width=0.5\textwidth]{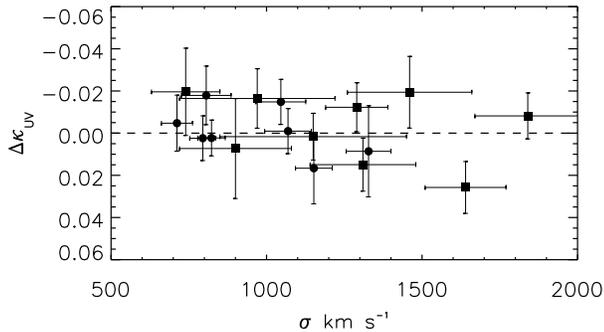}
\caption[]{The residuals about the models in Fig. \ref{fig:restev} plotted against velocity dispersion, $\sigma$. Filled circles and squares represent galaxy clusters with redshifts, $z<0.2$ and $z>0.2$ respectively, to demonstrate there is no redshift dependence. The MACS and LARCS cluster velocity dispersions are sourced from \cite{Ebeling2007} and \cite{LARCS2006} respectively. The correlation coefficient, r, for this plot is 0.27.}
\label{fig:ressiguv}
\end{figure}

We now investigate whether there are further trends in red sequence slope with other observable cluster properties to ensure that the result seen in \S\ref{sec:restf} is not due to a secondary correlation. The most obvious of these being the X-ray luminosity, which is a proxy for mass in a relaxed system. In Figs \ref{fig:reslxir} and \ref{fig:reslxuv} we plot the residual values of the slope about the model line in Fig. \ref{fig:evl} and fit in Fig. \ref{fig:restev} against $L_{X}$. From this we see no significant trend between scatter about the model and $L_{X}$ as the Pearson correlation coefficients, r , for the near-infrared and optical data are 0.17 and 0.24 respectively which is actually a weak anti-correlation between steepening negative red sequence slope and $L_{X}$. However, because of the magnitude of the errors on both the individual slope measurements and the models, we need to quantify what level of trend the errors could accommodate before becoming observable. For the near-infrared sample we find that a steepening greater than 0.010 in the slope in L$_{X}=26.6\times10^{44}$erg $s^{-1}$ can be ruled out with a 3 sigma confidence whereas for the rest frame $(U-V)$ sample we find this magnitude of change can occur in L$_{X}=12.3\times10^{44}$erg $s^{-1}$. The difference in mean L$_{X}$ between the LARCS and MACS samples is $8.5\times10^{44}$erg $s^{-1}$ for which we can rule out, with a 3 sigma confidence, increases in slope greater than 0.007, if we use the $(U-V)$ result, or 0.003 if we use the $J-K$ result. For the rest frame $\delta(U-V)/\delta V$ slope evolution in Fig. \ref{fig:restev} we see that the observed change in rest frame slope of $\sim0.04$ cannot be accounted for by the difference in $L_{X}$ alone and is, we believe, a definite trend with redshift. In Fig. \ref{fig:ressiguv} we also show that there is no significant trend between red sequence slope and another mass proxy, the cluster velocity dispersion, $\sigma$ (r$=0.27$), which is again a weak anti-correlation. As above we can rule out increases in slope greater than 0.010 with 3 sigma confidence between the average velocity dispersions of the MACS and LARCS samples.

\begin{figure}
\centering
\includegraphics[width=0.5\textwidth]{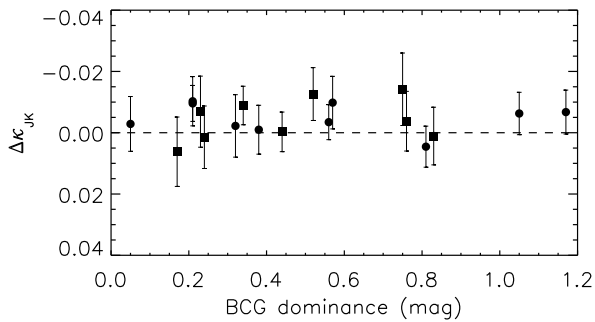}
\caption[]{The residuals about the models in Fig. \ref{fig:evl} plotted against BCG dominance. Filled circles and squares represent galaxy clusters with redshifts, $z<0.2$ and $z>0.2$ respectively, to demonstrate there is no redshift dependence. The correlation coefficient, r, for this plot is -0.09.}
\label{fig:resdomsigir}
\end{figure}

In addition to the X-ray luminosity and $\sigma$, we can look to the degree of BCG dominance as an indicator of the local environment within the cluster core. This parameterises the luminosity gap between the BCG and the next brightest galaxies on the red sequence and is defined as $\Delta m_{1-2,3}= (m_2+m_3)/2 -m_1$ where $m_1$ is the magnitude of the BCG and $m_2$ and $m_3$ are the magnitudes of the 2nd and 3rd brightest members respectively ~\citep{Kim2002, stott2007b}. The BCG may be the dominant elliptical in a cluster centre containing much smaller galaxies or it may be in a system were it is only marginally brighter than the next brightest members. In Fig. \ref{fig:resdomsigir} we demonstrate that there is no strong trend between red sequence slope and the degree of BCG dominance (r$=-0.09$) and can rule out increases of slope greater than 0.01 in 0.61 magnitudes of dominance with a 3 sigma confidence. 

The above results suggest that different cluster environments do not have a strong effect on the near-infrared or optical red sequence slope at a given redshift, demonstrating that the result in \S\ref{sec:sev} is robust. Previous studies within a similar $L_{X}$ range have also seen homogeneity in other related cluster properties such as the shape of the luminosity function and the blue galaxy fraction (\citealt{deprop1999}; \citealt{Wake2005}).

\section{Discussion}
\label{sec:sum}
In this work we have found a significant evolution in the rest frame slope of the red sequence in rich galaxy clusters between $z\sim0.5$ and $z\sim0.1$. We propose that this intrinsic evolution is due to galaxies falling into the cluster core and transforming onto the red sequence (\citealt{delucia2007dgr}; \citealt{Stott2007a}). If these galaxies have undergone recent star formation in filaments (e.g. \citealt{porter2008}), which has been quenched by interactions as they fall into the cluster, they will appear bluer than other passive galaxies in the cluster due to their young age. These galaxies will redden more rapidly with time than their old, luminous counterparts flattening the red sequence at progressively lower redshift. This relates to the concept of downsizing which describes the observation that star formation ceases in the largest galaxies first so that it takes place mainly in lower mass galaxies at late times \citep{Cowie1996}.

An alternative or additional explanation is that this result is caused by a differential chemical evolution along the red sequence. A differential chemical evolution along the sequence which preferentially enriches the faint galaxies could lead to a steeper red sequence slope at high redshift. The supposition of a metallicity differential along the sequence that decreases with redshift is borne out in the semi-analytic models as the gradient at $z=$0 is -0.0029$\pm$0.0004 Z$_{\odot}$ mag$^{-1}$ and at z=1 it is -0.061$\pm$0.009 Z$_{\odot}$ mag$^{-1}$. However, this may be non-physical with the primary goal of semi-analytic model is to match the observed properties of the local Universe and not necessarily the evolution from high redshift.

This result is in disagreement with the work of \cite{stanford98}, \cite{blakes2003} and \cite{mei2006} who find no evidence for intrinsic slope evolution. There are several potential reasons for this disagreement. We first note the scatter in the values of the slope which could obscure the intrinsic evolution in previous studies without the coverage per redshift bin we have in our sample. This scatter is important when choosing a low redshift comparison cluster, some earlier works use a single cluster, usually Coma, which we have shown to have an unusually steep red sequence compared to the mean of our low $z$ sample. The use of Coma would therefore tend to bias the  result towards no evolution. The choice of $(U - B)$ rest frame colour in previous studies means that they are observing a smaller dynamic range in colour than our $(U - V)$ sample. This is particularly important if the rest frame `$U-B$' colours do not effectively straddle the 4000\AA\, break (e.g. figure 4 of \citealt{mei2006}) which may result in contamination of the red sequence by both foreground galaxies and star forming cluster members. We also ensure that we perform all of the red sequence slope fits using the same homogenous method rather than importing slopes from other studies which may have a systematic offset. 

The good agreement found between our data and slope evolution models calculated from semi-analytical model of \cite{Bower2006} in Figures \ref{fig:evl} and \ref{fig:evr} is mainly due to the treatment of $K$ correction rather than any significant slope evolution. This demonstrates that although such models include `strangulation' of star-forming galaxies falling into cluster environments, they are unable to effectively reproduce the intrinsic evolution seen in our sample. This deficiency is also seen when studying the build up of the red sequence \citep{gilbank2008b}. We speculate that an improved  prescription for a strangulation-like process in the semi-analytic models will be an important factor in recreating observations of an intrinsic slope evolution and build up of the red sequence.

When looking at slope trends with other observables we see no relationship between cluster slope and X-ray luminosity, velocity dispersion or BCG degree of dominance. This suggests that there is very little variation between the red sequence slopes due to the different cluster environments demonstrating that the main result of this paper is robust to the differing global properties within our sample. This implies that searching for massive clusters using the colour magnitude relation (CMR), as employed by current and future large area optical/near-infrared surveys, is a viable method (e.g. RCS, \citealt{Glad2000}; Pan-STARSS; SDSS, \citealt{SDSS}; UKIDSS, \citealt{Law2007}, \citealt{Swin2007}).

\section{Acknowledgements}

We thank the referee for their useful comments which have improved the clarity and conclusions of this paper. Thanks also go to Richard Bower, Philip Best, Jim Geach and Matt Hilton for useful discussions. JPS acknowledges support through a Particle Physics and Astronomy Research Council and latterly a Science and Technology Facilities Council Studentship. KAP  acknowledges partial support from the Australian Research Council and partial support from a University of Queensland ResTeach Fellowship.

The United Kingdom Infrared Telescope is operated by the Joint Astronomy Centre on behalf of the Science and Technology Facilities Council of the U.K. We gratefully acknowledge the allocation of UKIRT service time for our observations.

This publication makes use of data products from the Two Micron All Sky Survey, which is a joint project of the University of Massachusetts and the Infrared Processing and Analysis Centre/California Institute of Technology, funded by the National Aeronautics and Space Administration and the National Science Foundation.

The Millennium Simulation databases used in this paper and the web application providing online access to them were constructed as part of the activities of the German Astrophysical Virtual Observatory.

\bibliographystyle{mn2e}
\bibliography{slope}

\end{document}